\documentstyle[epsfig]{mn}

\newif\ifAMStwofonts

\newcommand{\Halpha}{H$\alpha$}
\newcommand{\halpha}{H$\alpha$}
\newcommand{\Hbeta}{H$\beta$}
\newcommand{\Hgamma}{H$\gamma$}
\newcommand{\novacyg}{V404~Cyg}
\newcommand{\novamon}{A\,0620--00}
\newcommand{\ergspersec}{erg\,s$^{-1}$}
\newcommand{\ergspercmsec}{erg\,cm$^{-2}$\,s$^{-1}$}

\ifoldfss
  \ifCUPmtlplainloaded \else
    \NewTextAlphabet{textbfit} {cmbxti10} {}
    \NewTextAlphabet{textbfss} {cmssbx10} {}
    \NewMathAlphabet{mathbfit} {cmbxti10} {} 
    \NewMathAlphabet{mathbfss} {cmssbx10} {} 
  \fi
  \ifAMStwofonts
    \ifCUPmtlplainloaded \else
      \NewSymbolFont{upmath} {eurm10}
      \NewSymbolFont{AMSa} {msam10}
      \NewMathSymbol{\upi}     {0}{upmath}{19}
      \NewMathSymbol{\umu}     {0}{upmath}{16}
      \NewMathSymbol{\upartial}{0}{upmath}{40}
      \NewMathSymbol{\leqslant}{3}{AMSa}{36}
      \NewMathSymbol{\geqslant}{3}{AMSa}{3E}

    \fi
  \fi
\fi 

\ifnfssone
  \newmathalphabet{\mathit}
  \addtoversion{normal}{\mathit}{cmr}{m}{it}
  \addtoversion{bold}{\mathit}{cmr}{bx}{it}

    \selectfont #1}}
  \newmathalphabet{\mathbfit} 
  \addtoversion{normal}{\mathbfit}{cmr}{bx}{it}
  \addtoversion{bold}{\mathbfit}{cmr}{bx}{it}
  \newmathalphabet{\mathbfss} 
  \addtoversion{normal}{\mathbfss}{cmss}{bx}{n}
  \addtoversion{bold}{\mathbfss}{cmss}{bx}{n}
  \ifAMStwofonts
    \ifCUPmtlplainloaded \else
      \UseAMStwoboldmath
      \makeatletter
      \new@mathgroup\upmath@group
      \define@mathgroup\mv@normal\upmath@group{eur}{m}{n}
      \define@mathgroup\mv@bold\upmath@group{eur}{b}{n}
      \edef\UPM{\hexnumber\upmath@group}
      \new@mathgroup\amsa@group
      \define@mathgroup\mv@normal\amsa@group{msa}{m}{n}
      \define@mathgroup\mv@bold\amsa@group{msa}{m}{n}
      \edef\AMSa{\hexnumber\amsa@group}
      \makeatother
      \mathchardef\upi="0\UPM19
      \mathchardef\umu="0\UPM16
      \mathchardef\upartial="0\UPM40
      \mathchardef\leqslant="3\AMSa36
      \mathchardef\geqslant="3\AMSa3E
    \fi
  \fi
\fi 

\ifnfsstwo
  \DeclareMathAlphabet{\mathbfit}{OT1}{cmr}{bx}{it}
  \SetMathAlphabet\mathbfit{bold}{OT1}{cmr}{bx}{it}
  \DeclareMathAlphabet{\mathbfss}{OT1}{cmss}{bx}{n}
  \SetMathAlphabet\mathbfss{bold}{OT1}{cmss}{bx}{n}
  \ifAMStwofonts
    \ifCUPmtlplainloaded \else
      \DeclareSymbolFont{UPM}{U}{eur}{m}{n}
      \SetSymbolFont{UPM}{bold}{U}{eur}{b}{n}
      \DeclareSymbolFont{AMSa}{U}{msa}{m}{n}
      \DeclareMathSymbol{\upi}{0}{UPM}{"19}
      \DeclareMathSymbol{\upartial}{0}{UPM}{"40}
      \DeclareMathSymbol{\leqslant}{3}{AMSa}{"36}
      \DeclareMathSymbol{\geqslant}{3}{AMSa}{"3E}
    \fi
  \fi
\fi 

\ifCUPmtlplainloaded \else
  \ifAMStwofonts \else 
    \def\upi{\pi}
    \def\umu{\mu}
    \def\upartial{\partial}
  \fi
\fi

\title[H$\alpha$ Flares from V404~Cyg in Quiescence] 
{H$\alpha$ Flares from V404~Cyg in Quiescence}
\author[R. I. Hynes et al.]
       {R. I. Hynes$^1$\thanks{e-mail: rih@astro.soton.ac.uk}, 
	C. Zurita$^2$,
	C. A. Haswell$^3$,
        J. Casares$^2$,
	P. A. Charles$^1$, \newauthor 
	E. P. Pavlenko$^4$,
	S. Yu. Shugarov$^5$,
	D. A. Lott$^3$\\
$^1$Department of Physics and Astronomy, University of Southampton, 
    Southampton, SO17 1BJ\\
$^2$Instituto de Astrof\'\i{}sica de Canarias, 38200 La Laguna,
    Tenerife, Spain\\
$^3$Department of Physics and Astronomy, The Open University, Walton
    Hall, Milton Keynes, MK7 6AA\\
$^4$Crimean Astrophysical Observatory, Nauchny, 98409 Crimea, Ukraine\\
$^5$Sternberg Astronomical Institute, Moscow State 
University, Universitetskii pr.13, Moscow 119899, Russia\\
} 
\date{Accepted ?.
      Received ?;
      in original form ?}

\pagerange{\pageref{firstpage}--\pageref{lastpage}}
\pubyear{2001}

\begin{document}
\maketitle
%
%
\begin{abstract}
We present a spectrophotometric study of short-term optical
variability in the quiescent black hole X-ray transient \novacyg.
This includes two nights of high time-resolution \Halpha\ spectroscopy
with which we resolve much of the time-variability, and a further six
nights of archival spectroscopy with lower time-resolution but higher
spectral-resolution.  We find significant variability in most of the
data considered, with both the \Halpha\ line and the continuum often
varying in a correlated way.  This includes both dramatic flares
lasting a few hours in which the line flux nearly doubles and
lower-level flickering.  The strongest flares involve development of
asymmetry in the line profile, with the red wing usually strongest
independent of orbital phase.  It is unclear why this is the case, but
we discuss several possible explanations.
We consider the energetics of the flares and compare with plausible
models including chromospheric activity on the companion star, local
magnetic reconnection events within the disc and a varying irradiation
from close to the black hole.  
Based on the line profile changes during the flares, we conclude that
the most likely origin for the variability is variable photoionisation
by the central source, although local flares within the disc cannot be
ruled out.
\end{abstract}
%
%
%
%
\section{Introduction}
\label{IntroSection}
Quiescent black hole X-ray transients (BHXRTs) provide the best
evidence we have for the existence of stellar mass black holes.  These
X-ray binaries contain a relatively low-mass star (usually, but not
always, $<1$\,M$_{\odot}$) accreting onto a likely black hole via
Roche lobe overflow and an accretion disc.  In their quiescent state
the accretion flow becomes extremely faint and so the companion star
can be directly observed.  It is the companion, and in particular its
radial velocity variations, which provide the key to measuring the
black hole mass (Charles 1998).

One of the most secure cases for a black hole is for \novacyg, which
contains a compact object with a mass function of
$6.08\pm0.06$\,M$_{\odot}$ (Casares, Charles \& Naylor 1992), twice
the usually accepted upper limit to a neutron star mass (Rhoades \&
Ruffini 1974).  The best estimate for the actual mass is
12$\pm$2\,M$_{\odot}$ (Shahbaz et al.\ 1994) and this system likely
contains the most massive stellar mass black hole for which we have
dynamical evidence.  It has the longest orbital period of the BHXRTs,
at 6.5\,days, with a K0\,IV companion star (Casares et al.\ 1993).
\novacyg\ is the most luminous of these systems in quiescence with
$L_{\rm X} \sim 10^{33}-10^{34}$\,erg\,s$^{-1}$, $\sim1000\times$
brighter than the prototypical BHXRT, \novamon\ (Garcia et al.\ 2001).
These extremes are unlikely to be coincidental and the high $L_{\rm
X}$ may be a consequence of the long orbital period, the evolved donor
star, and/or the high mass of the black hole.

The high quiescent $L_{\rm X}$ indicates that even in its quiescent
state, \novacyg\ is not completely dormant, but that some accretion
continues.  This is also indicated by the substantial short-term
(sub-orbital) variability that \novacyg\ exhibits.  In the optical
this has an amplitude of 0.1--0.2\,mag (Wagner et al.\ 1992), at least
partly due to a 6\,hr quasi-periodic oscillation (Casares et al.\
1993; Pavlenko et al.\ 1996).  Short timescale infrared variations
also appear to be present (Sanwal et al.\ 1996) and the \Halpha\ line
profiles change significantly on short timescales (Casares \& Charles
1992; Casares et al.\ 1993).  At X-ray energies, Wagner et al.\ (1994)
saw changes of a factor of 10 on timescales $<0.5$\,day with {\it
ROSAT} and more recently Garcia et al.\ (2000) presented a {\it
Chandra} lightcurve exhibiting a factor of $\sim 2$ variability in a
few ksec.  Finally \novacyg\ remains a variable radio source on
timescales of days: Hjellming et al.\ (2000) reported typical
variability in flux of 0.1-0.8\,mJy with a maximum quiescent flux of
1.5\,mJy.

\novacyg\ is not the only BHXRT to exhibit variability in quiescence.
Haswell (1992) discovered significant optical flaring in \novamon\ on
timescales of minutes.  Zurita et al.\ (in preparation) have found
optical photometric variability in other systems: in addition to
\novacyg\ and \novamon, they also identify strong variability in 
GRO\,J0422+32 and also in the neutron star system Cen X-4.  This
kind of behaviour thus seems to be common in quiescent BHXRTs.

\begin{table*}
\caption{Log of spectroscopic and photometric observations of
\novacyg\ from 1992 (La Palma) and 1999 (La Palma, CrAO and SAI).}
\label{ObsTable}
\begin{center}
\begin{tabular}{llccccc}
\hline
Date& Telescope& UT Range& Exposures& Number of& Wavelength & Resolution\\
    &          &         & Time (s) & Exposures& range (\AA)& (\AA)      \\
\noalign{\smallskip}
{\it La Palma Spectroscopy} &       &      &   &            &     \\
1992 Jul 5  & WHT & 00:51--04:43 & 1800 & 8   & 6200--6600 & 0.5 \\
1992 Jul 6  & WHT & 00:38--05:04 & 1800 & 8   & 6200--6600 & 0.5 \\
1992 Jul 8  & WHT & 00:33--04:58 & 1800 & 10  & 6200--6600 & 0.5 \\
1992 Jul 9  & WHT & 00:56--04:24 & 1800 & 7  & 6200--6600 & 0.5 \\
1992 Jul 10 & WHT & 00:20--04:53 & 1800 & 9  & 6200--6600 & 0.5 \\
1992 Jul 11 & WHT & 23:46--03:43 & 1800 & 8  & 6200--6600 & 0.5 \\
\noalign{\smallskip}
1999 Jul 6  & WHT & 21:27--04:02 & 180 & 98 & 6000--7400 & 4 \\
1999 Jul 7  & WHT & 21:30--04:04 & 240 & 83 & 6000--7400 & 4 \\
\noalign{\smallskip}
{\it La Palma Photometry} &       &      &   &            &     \\
1992 Jul 5  & JKT & 01:09--05:15 & 200 & 33  &&\\
1992 Jul 6  & JKT & 00:34--05:17 & 200 & 50  &&\\
1992 Jul 8  & JKT & 00:28--05:11 & 200 & 63  &&\\
1992 Jul 9  & JKT & 00:56--04:31 & 200 & 44  &&\\
1992 Jul 10 & JKT & 00:13--04:40 & 200 & 49  &&\\
1992 Jul 11 & JKT & 23:40--04:10 & 200 & 45  &&\\
\noalign{\smallskip}
1999 Jul 6  & JKT & 22:03--03:56 & 60  & 205  &&\\
1999 Jul 7  & JKT & 21:54--03:57 & 60  & 259  &&\\
\noalign{\smallskip}
{\it CrAO and SAI Photometry} &       &      &   &            &     \\
1999 Sep 11 &  38\,cm & 20:53--00:45 & 120 & 84  &&\\
1999 Sep 12 &  38\,cm & 19:29--01:11 & 200 & 70  &&\\
1999 Sep 16 &  38\,cm & 20:36--21:58 & 180 & 23  &&\\
1999 Sep 17 &  38\,cm & 20:12--00:01 & 180 & 43  &&\\
1999 Sep 18 &  38\,cm & 19:26--20:24 & 180 & 11  &&\\
1999 Sep 21 & 125\,cm & 22:31--01:34 &  90 & 96  &&\\
1999 Sep 22 & 125\,cm & 22:43--00:45 &  90 & 56  &&\\
1999 Sep 23 & 125\,cm & 23:26--02:09 &  90 & 73  &&\\
1999 Oct 16 &  38\,cm & 20:52--22:04 & 180 & 23  &&\\
1999 Nov 05 &  38\,cm & 20:15--21:07 & 200 & 11  &&\\
1999 Nov 11 & 125\,cm & 19:21--19:26 &  20 &  2  &&\\
1999 Nov 14 & 125\,cm & 17:59--18:14 &  90 &  7  &&\\
1999 Nov 19 &  60\,cm & 18:28--18:33 & 120 &  3  &&\\
\hline
\end{tabular}
\end{center}
\end{table*}

It is important to understand these variations as the properties of
the variability can yield important clues about the nature of the
quiescent accretion flow and provide an observational test of models
for this flow.  For example, advection dominated accretion flow (ADAF)
models for V404 Cyg have predicted that the non-stellar optical
continuum is dominated by synchrotron emission from very close to the
black hole (Narayan, Barret \& McClintock 1997; Quataert \& Narayan
1999).  One would most naturally expect the ADAF temporal variability
to be different to that of the emission lines which show a double
peaked profile, indicative of an origin in the outer disc, although
there is also a possibility that the lines could be photoionised by
the ADAF.  Alternatively, both line and non-stellar continuum emission
could originate in the outer disc, with variability likely associated
with local magnetic reconnection events.  One would expect in this
case rapid and kinematically localized enhancements in the line
emission, although the flares might be broadened by motions within the
flaring material.  Another important source of variability may be the
companion star.  This is a K0\,IV star, tidally locked in a 6.5\,day
orbit and likely similar to active stars in RS CVn binaries.

To investigate these issues, in 1999 July we obtained optical
spectroscopy at high time-resolution so as to reveal the spectral
signature of the short-term variability.  We obtained simultaneous
photometry to ensure a reliable flux calibration; hence we can
accurately study both line and continuum variations.  The low spectral
resolution also provides us with kinematic resolution of emission line
variations.  To place these observations in a broader context we
include longer term photometric monitoring observations taken during
1999.  We also re-analysed high spectral resolution data from a
spectrophotometric campaign on \novacyg\ performed in 1992 July.  A
preliminary analysis of the latter data appeared in Zurita, Casares \&
Charles (2000).  We describe the observations in
Section~\ref{ReductionSection}.  Section~\ref{SpecSection} presents
the average spectrum of \novacyg\ from 1999 together with that of the
nearby blended star.  In Section~\ref{LightcurveSection} we compare
our line and continuum lightcurves.  Section~\ref{SpecVarSection}
examines the spectrum of the variability in more detail and in
Section~\ref{DiscussionSection} we discuss our results in the light of
current models for quiescent BHXRTs.
%
%
\section{Observations}
\label{ReductionSection}
We observed V404 Cyg from the Observatorio del Roque de los Muchachos
on La Palma on 1992 July 5--6 and 8--11 and 1999 July 6--8 using the
1.0\,m Jacobus Kapteyn Telescope (JKT) for photometry and the 4.2\,m
William Herschel Telescope (WHT) for spectroscopy.  Conditions on all
nights were good, with little or no dust or cirrus and seeing
typically $\sim1$\,arcsec.  Further photometry was obtained with the
0.38\,m telescope of the Crimean Astrophysical Observatory (CrAO) and
the 1.25\,m and 0.6\,m telescopes of the Sternberg Astronomical
Institute (SAI)  from 1999 September 11 to November 19.  A full log
of observations used is given in Table~\ref{ObsTable}.

%
\subsection{JKT Photometry}
To ensure accurate photometric calibration of the WHT spectroscopy, we
obtained simultaneous $R$ band photometry.  In 1999 we used the SITe2
CCD camera on the JKT.  We used integration times of 60\,s on both
nights, obtaining a temporal resolution of $\sim$80\,s.  All images
were corrected for bias and flat-fielded in the standard way using
{\sc iraf}.  Although the seeing conditions were good, we were not
able to cleanly separate the contribution of the target and its nearby
contaminating star so we performed simple aperture photometry around
both stars.  The contribution of the blended star, assumed constant,
was then subtracted assuming $R=17.52$ (Casares et al.\ 1993).  We
used an aperture of 9\,pixel radius ($\sim$3\,arcsec) for our object
and several comparison stars within the field of view, which were
checked for variability during each night and during the entire data
set.  We also explored any correlation between the target differential
variations and the seeing, but found none.  The 1992 data were
obtained with the GEC CCD camera on the same telescope. The typical
integration time was 200\,s. Aperture photometry was performed with an
aperture of 11\,pixel radius ($\sim$3.6\,arcsec). Further details of
these observations can be found in Pavlenko et al.\ (1996).
\subsection{WHT Spectroscopy}
\label{SpecRedSection}
We obtained the 1999 time-resolved spectrophotometry using the ISIS
dual-arm spectrograph on the WHT.  To maximise efficiency and minimise
readout time and noise, we used the single red-arm mode with the
relatively low-resolution R316R grating and the TEK4 CCD.  Cycle times
were 203\,s on 1999 July 6--7 and 263\,s on 1999 July 7--8.  A wide
slit (4\,arcsec) was used to maximise photometric accuracy, so our
spectral resolution was determined by the seeing ($\sim1.0$\,arcsec),
and was typically $\sim4$\,\AA.  Bias correction and flat fielding
were performed using standard {\sc iraf} techniques.

Observations of \novacyg\ are made difficult by the presence of a
fainter star 1.5\,arcsec to the north and usually blended with it
(Udalski \& Kaluzny 1991; Casares et al.\ 1993).  Because of this, we
chose a slit alignment passing through the line of centres of the two
stars.  Fortunately this also coincided with a third, isolated,
relatively bright star.  Hence we could use the isolated star to
define the spatial profile along the slit, and then deblend the
spectra of V404 Cyg and the nearby star, following the method
summarised in Hynes et al. (1999) and described in more detail by
Hynes (1999,2001).  For these data, we used the isolated star to
construct a wavelength independent semi-analytical spatial profile for
each spectrum.  This comprised a Voigt profile with a numerical
correction determined for the profile wings.  The profile deblending
optimal extraction method does reject most cosmic rays and bad pixels,
but a few remained; these were corrected by hand.

Wavelength calibration was interpolated between contemporaneous
exposures of a copper-neon/copper-argon arc lamp.  The dominant
residual error was caused by the uncertain positioning of the star
within the wide slit.  The sky lines could not be used to correct the
wavelength solution because these fill the slit and the stars
positioning relative to these is also uncertain; hence we must use
spectral features of the objects.  Unfortunately the template star
shows no strong spectral features in this region.  We experimented
with using both the \Halpha\ absorption feature in the faint blended
star and various stellar absorption features in V404 Cyg (over
$\lambda\lambda$6380--6520\,\AA) and found that the latter gave less
scatter in the corrections, which were calculated by cross-correlating
the individual spectra in this region with an average.  Of course the
photospheric features do move, so this correction represents a
transformation into the companion rest frame.  We then use the known
radial velocity curve of Casares \& Charles (1994) to transform
into an inertial frame.  Initial flux calibration was done relative to
the template star.
This was in turn calibrated relative to the spectrophotometric
standard Cyg OB2, No.\ 9 (Massey et al.\ 1988).  Finally we performed
synthetic $R$ band photometry of the sum of calibrated spectra of
\novacyg\ and the blended star and
slit losses were corrected
by comparing with the simultaneous JKT photometry.

\novacyg\ was also observed in the \Halpha\ region with the ISIS
spectrograph on the WHT on the nights 1992 July 5--6 and 8--11. A
total of 50 spectra were obtained using individual exposures times of
1800\,s. The main aim of these observations was to measure the
rotational broadening of the companion star, so a narrow (0.8\,arcsec)
slit was used in combination with a 1200 line mm$^{-1}$ grating to
yield a high spectral resolution of 0.5\,\AA\ (FWHM). The wavelength
range covered was $\sim\lambda\lambda$6200--6600\,\AA.  The spectra
were calibrated using the flux standard BD$+17^{\circ}4708$ (Oke \&
Gunn 1983).  Further details of these observations can be found in
Casares \& Charles (1994). As the slit was narrower, the slit loss
correction is more dramatic than in the 1999 observations. We made the
correction, as is explained in the first part of this section, by
comparing the simultaneous JKT photometric fluxes and the
spectroscopic fluxes.
\subsection{Contemporaneous Photometry}
As part of an ongoing campaign to monitor \novacyg\ in quiescence,
observations were carried out at the 0.38\,m Cassegrain telescope of
the Crimean Astrophysical Observatory (CrAO) and the 1.25\,m and
0.6\,m telescopes of the Sternberg Astronomical Institute (SAI) in a
broad $R+I$ system, using SBIG ST-7 cameras. The images were
dark-subtracted, flat-fielded and analysed with a profile/aperture
photometry package developed by Goransky (SAI).
%
%
\section{Average spectra}
\label{SpecSection}

\begin{figure}
\begin{center}
\epsfig{angle=90,width=3.4in,file=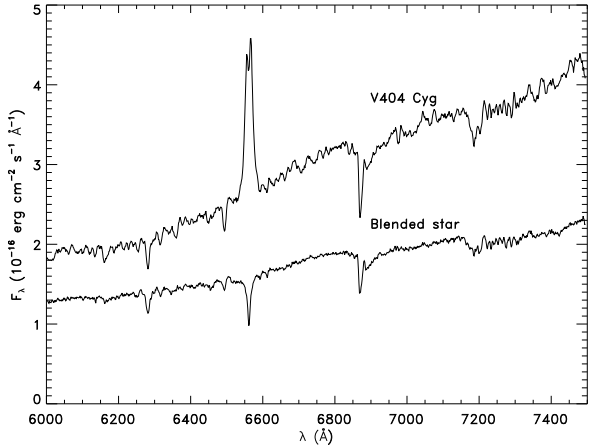}
\caption{Deblended spectra of \novacyg\ and the blended star.  All the
spectra from 1999 July have been averaged.  The deblending
algorithm has clearly separated the spectra cleanly.  Note the
contrast between \Halpha\ emission in \novacyg\ and absorption in the
blended star.}
\label{SpecFig}
\end{center}
\end{figure}

In Fig.~\ref{SpecFig} we show the average spectra of \novacyg\ and the
nearby blended star from 1999.  To our knowledge, this is the first
time that the spectra of the two stars have been separated; most
spectroscopic observations either yield a combined spectrum, or align
the slit perpendicular to the line of centres to minimise the
contribution from the blended star.  Contamination by this star will
distort the spectrum and must be considered (e.g.\ Shahbaz et al.\
1996) so it is worthwhile to present the separated spectra for
comparison.

Photometry of the blended star yielded $B-V=1.69$, $V-R=1.38$ (Casares
et al.\ 1993).  For different trial reddenings, the best agreement
with colours of main sequence stars comes for a K4--5 star with
$E(B-V)\sim0.5$.  The reddening line in a $B-V$ vs.\ $V-R$ plot is
almost parallel to the main-sequence for spectral types earlier than
K, however, so more reddened earlier types are also possible.  The
only strong feature visible in the blended star's spectrum is \Halpha\
absorption.  This is stronger than expected for a late K star and
instead favours an F type identification; a higher resolution spectrum
would be needed for a more precise classification.  A weak 6495\,\AA\
absorption line also appears to be present; this feature can be
present in F stars (see e.g.\ the templates used by Orosz \& Bailyn
1997), so is consistent with this interpretation.  For the photometric
colours to agree with an F star requires $E(B-V)\sim 1.3\pm0.1$, or
$A_V \sim 4.0\pm0.3$.  This implies a dereddened brightness of
$V\sim14.9$ and a distance of $2.0\pm0.5$\,kpc for a main sequence
star.  Both reddening and distance are similar to the values for
\novacyg\ (Casares et al.\ 1993) so this is a plausible solution.
Further support for a similar reddening to \novacyg\ is the presence
of the 6284\,\AA\ diffuse interstellar band with a similar strength in
both objects.  If the blended star is an F star then we expect it to
have no significant CO features in the infrared.  Consequently the
infrared disc contribution will be reduced further below the 14\,per
cent upper limit of Shahbaz et al.\ (1996).

For the following analysis we will require the \Halpha\ line
luminosity.  We observe an average line flux of
$4.8\times10^{-15}$\,\ergspercmsec, and Casares et al.\ (1993)
estimate an optical extinction of $A_V=4.0$.  Assuming the gas to dust
scaling of Bohlin, Savage \& Drake (1978), this extinction corresponds
to a neutral hydrogen column of $\sim7\times10^{21}$\,cm$^{-2}$, which
is also consistent with that inferred from the {\it Chandra} spectra
(Garcia et al.\ 2000).  With $A_V=4.0$ and an average Galactic
extinction curve (Seaton 1979) we infer an interstellar attenuation at
\Halpha\ of a factor of $\sim20$.  A distance of 3.5\,kpc was adopted
by Garcia et al.\ (2001) and to facilitate comparison with their X-ray
results we adopt the same value.  Finally, to convert observed fluxes
into luminosities we also assume isotropic emission.  This is probably
not correct for emission from a disc, but as V404 Cyg appears to be an
intermediate inclination system (Shahbaz et al.\ 1994), the difference
in the ratio of observed flux to luminosity between isotropic and disc
emission is likely to be small, and this is acceptable for a crude
approximation.  We finally derive an average, dereddened \Halpha\
luminosity of $1.4\times10^{32}$\,\ergspersec.
%
%
\begin{figure}
\begin{center}
\epsfig{angle=90,width=3.4in,file=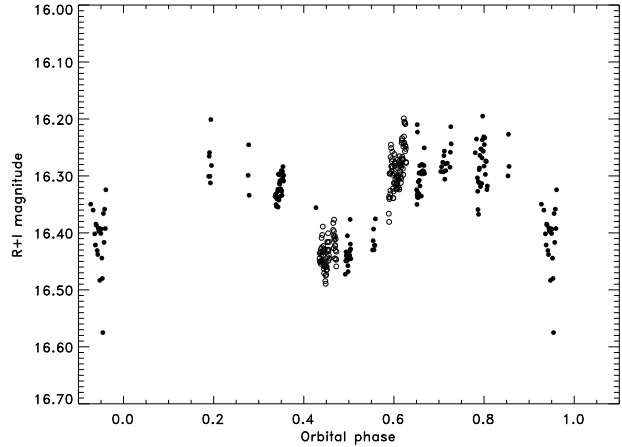}
\caption{Phase-folded photometric lightcurves from CrAO and SAI (solid
circles) and the JKT (open circles).  All lightcurves have been binned
in groups of three to improve signal to noise, and the $R$ band JKT
points have been shifted upwards by 0.38\,mag to bring them into
approximate agreement with the $R+I$ band observations.}
\label{PavlenkoFig}
\end{center}
\end{figure}

\begin{figure*}
\begin{center}
\epsfig{width=3.45in,file=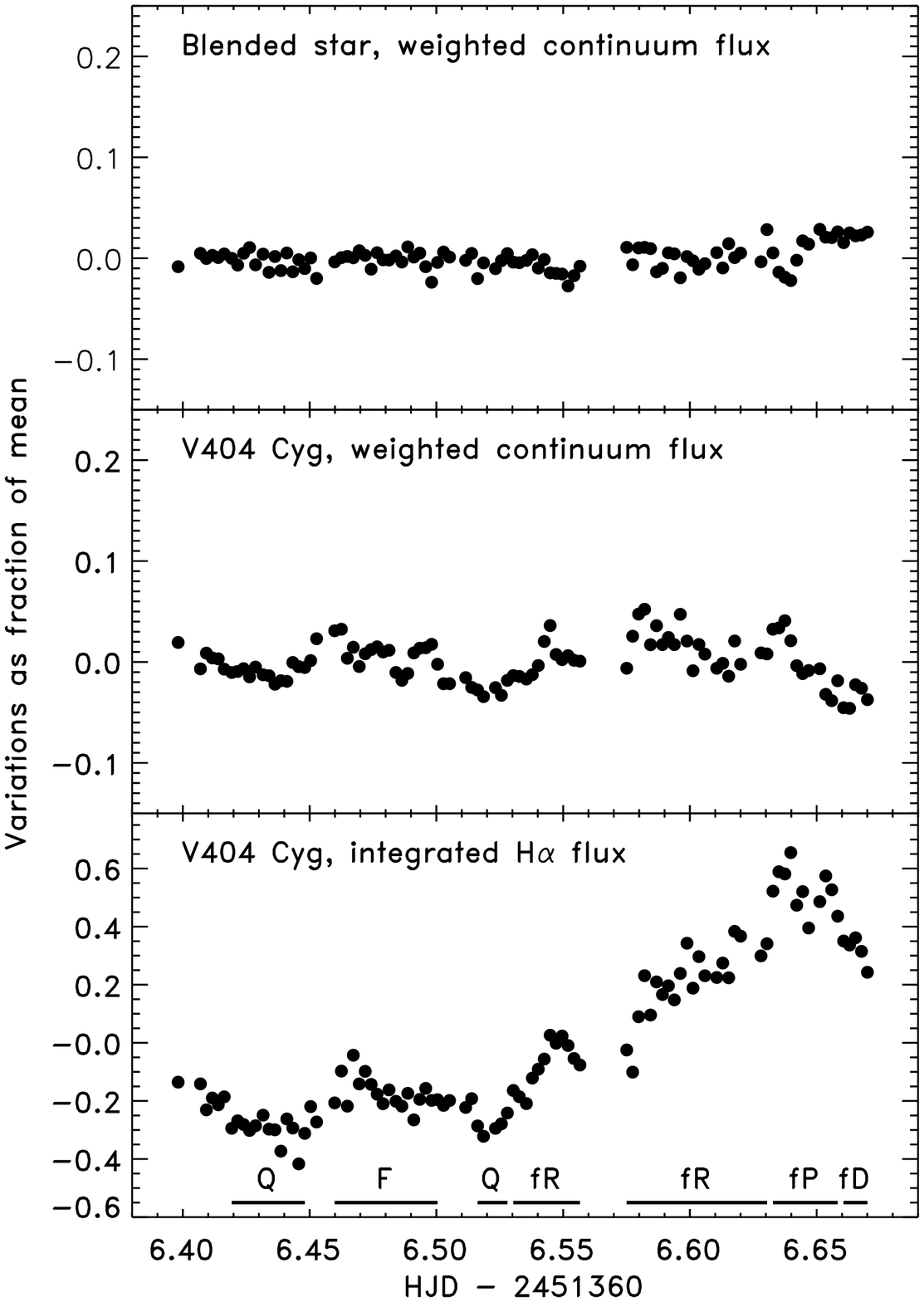}
\epsfig{width=3.45in,file=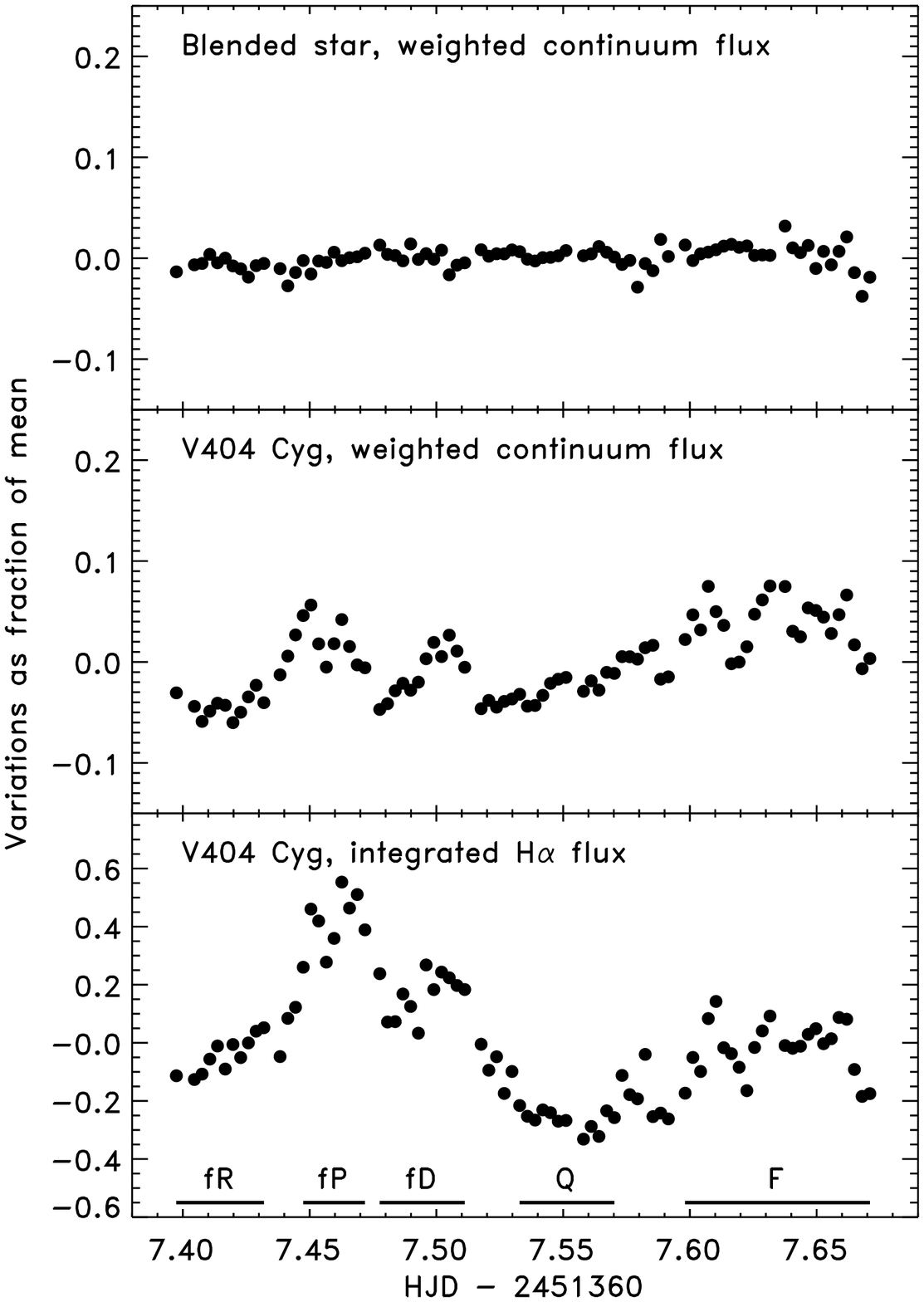}
\caption{Continuum light curves of \novacyg\ and the nearby blended
star, and \Halpha\ emission light curve of \novacyg\ from the 1999
spectrophotometry on July 6--7 (left) and 7--8 (right).  Continuum
measurements are based on a weighted average across the spectral range
covered, with lines masked out and with higher weight given where the
signal-to-noise ratio is highest.  Note that the vertical range of the
\Halpha\ plot is much larger than for the continuum light curves.  The
blended star is fainter than V404~Cyg, so provides an upper limit to
the statistical and systematic errors expected in the lightcurve of
\Halpha.  The notation for dividing up the lightcurve is Q: `quiescent
state', F: isolated flickering regions, somewhat brighter than Q and
fR, fP and fD denote the flare rise, peak and decline respectively.}
\label{LCFig}
\end{center}
\end{figure*}

\section{Lightcurves}
\label{LightcurveSection}
\subsection{The orbital lightcurve}
In Fig.\ \ref{PavlenkoFig} we present phase-folded photometric
lightcurves from CrAO, SAI and the JKT in 1999.  Scales have been
chosen to facilitate comparison with Fig.\ 1 of Pavlenko et al.\
(1996).  As with the 1992 data described there, the lightcurve is
dominated by the double-humped ellipsoidal modulation, but superposed
upon it are shorter term variations visible within a single night.
The 1999 data were obtained in a redder bandpass ($R+I$) than those
obtained in 1992, so are not directly comparable, but it does appear
that the amplitude of both ellipsoidal and short-term variations is
less.  Apparent changes in the ellipsoidal lightcurve of \novacyg\
based on the monitoring campaign have already been noted by Pavlenko
et al.\ (2001).

In the earlier data the short term variations appeared quasi-periodic
(Pavlenko et al.\ 1996).  We performed a period search on the 1999
data to test if this was still the case.  Before doing the analysis we
removed the 6.4714\,d orbital modulation and its first two harmonics.
The periodogram of the residuals showed several peaks and it is
difficult to choose between them, or even prove that the variations
are quasi-periodic.  Formally the peak with the highest significance
has period $(0.25404\pm0.00005)$\,d, $\sim6.1$\,hr with mean
amplitude $\sim0.1$\,mag.  This is close to the period found in the
earlier data, but other periods are possible and the case is
inconclusive.  


\subsection{The high temporal resolution spectra}

Light curves from the 1999 high temporal resolution spectra are shown
in Fig.~\ref{LCFig}.  On both nights we find the variability is
dominated by a single large flare, although the profiles of these
flares are complex.  On the second night the flare is clearly seen in
the continuum as well, with the same structure.  The amplitude of the
line flare is much larger than that in the continuum, but this is
likely because the continuum is heavily diluted by light from the
mass donor star; Casares et al.\ (1993) estimate 90\,percent of the
light in the $R$ band is from the mass donor.  After correction for
this, the fractional amplitudes of the flares in line and inferred
accretion flow continuum are similar.  On the first night, the
continuum does not reflect the large flare, although finer structure
does appear common to the lightcurves and a weak continuum flare does
coincide with the peak of the \Halpha\ flare.

\begin{figure}
\begin{center}
\epsfig{angle=90,width=3.4in,file=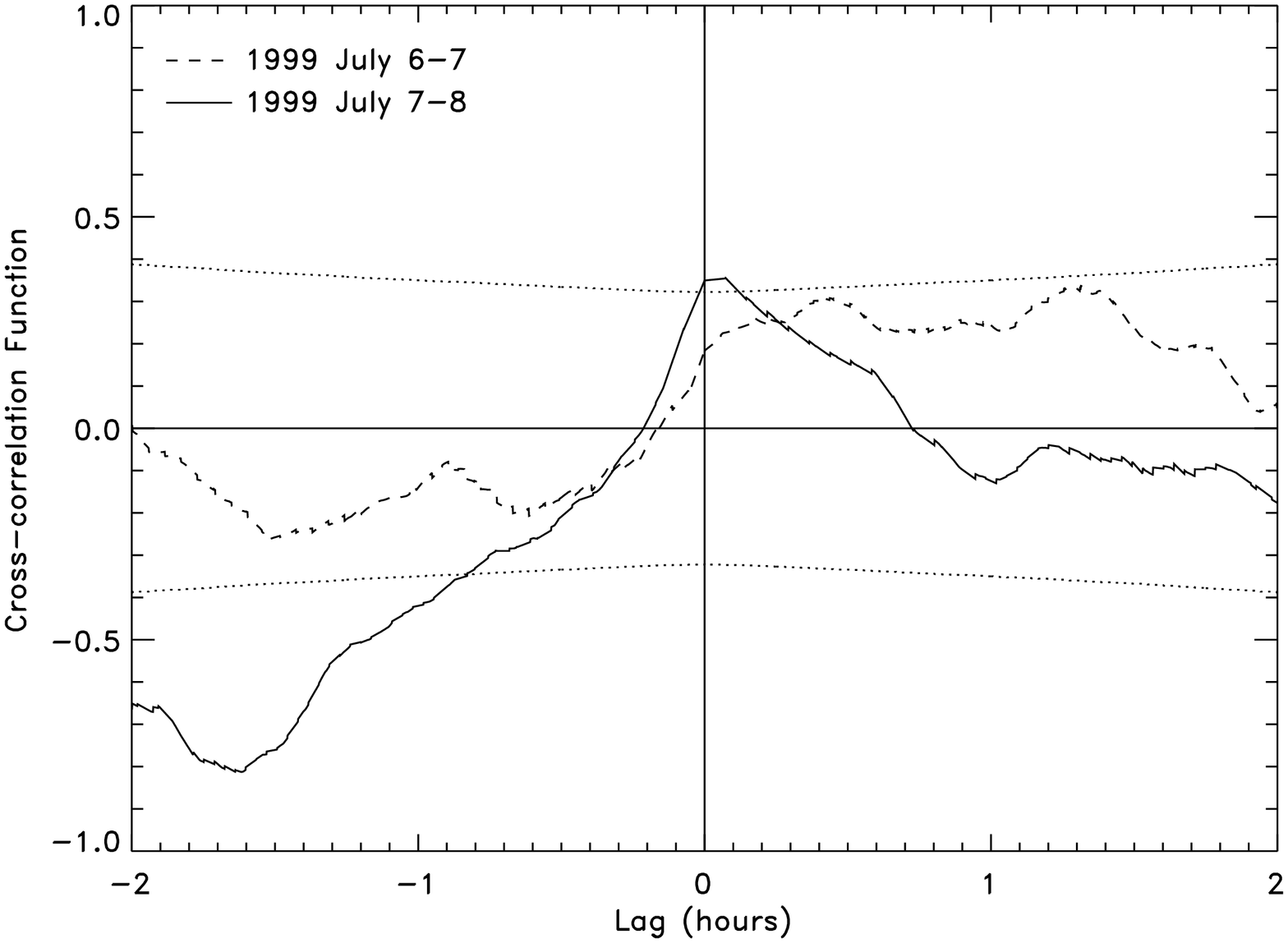}
\epsfig{angle=90,width=3.4in,file=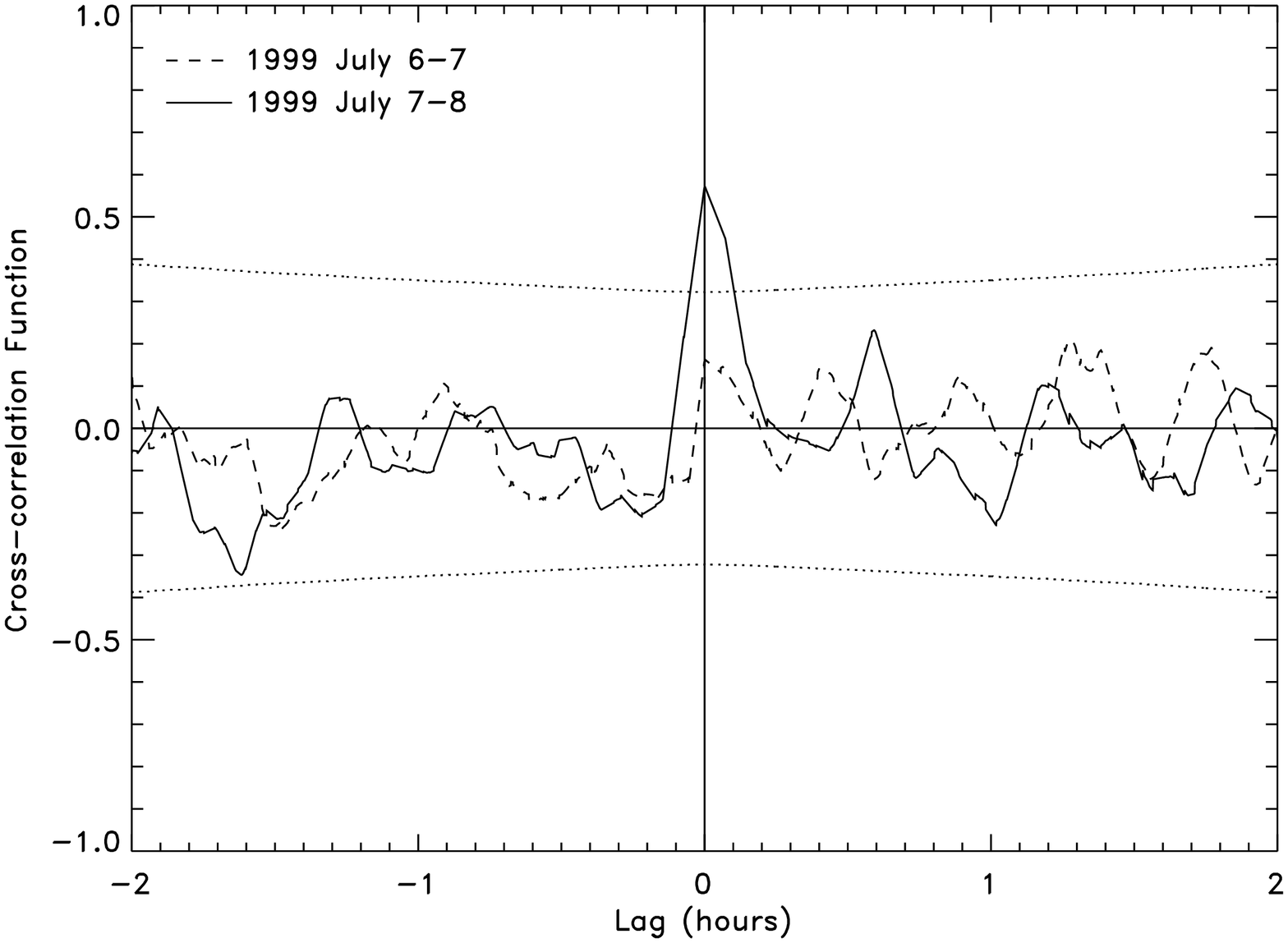}
\caption{Cross-correlation functions between line and continuum from
1999 data.  A positive lag would correspond to line variations lagging
behind those in the continuum.  Dotted lines indicate 3\,$\sigma$
limits on expected coincidental correlations (almost the same on the
two nights).  Upper panel shows the cross-correlations of the raw
light curves; lower panel is after applying a high-pass Fourier filter
with a cut-off at 1\,hr$^{-1}$.}
\label{CCFFig}
\end{center}
\end{figure}

To test if the flaring is simultaneous, or if one lags the other, we
calculated cross-correlation functions between them.  The lightcurves
are reasonably well sampled so we use the interpolation cross
correlation function (Gaskell \& Peterson 1987; White \& Peterson
1994).  We also calculated discrete correlation functions (Edelson \&
Krolik 1988) and the results were almost identical.
Our CCFs for the two nights are shown in Fig.~\ref{CCFFig}.  As noted
above, the correlation between line and continuum does not show up
well on the first night, and at 3\,$\sigma$ confidence is not
significant.  On the second night a clear peak is seen, significant at
the 3\,$\sigma$ level.  The peak shows no large lag between line and
continuum, but does appear asymmetric.  This could correspond to line
variations lasting longer than those in the continuum, perhaps due to
a significant recombination timescale.  The lack of correlation on the
first night, and the strange structure of the CCF on the second night
are at least partly due to low-frequency variations, in particular the
absence of large continuum flares when large \Halpha\ flares occur.
On the first night there is no prominent continuum flare, but some
shorter timescale variations do appear correlated.  On the second
night only the peaks of the large flare are well reproduced by the
continuum, not the broader base.  To further explore this, we removed
the poorly correlated low-frequency variations by applying a high-pass
Fourier filter to the lightcurves with a cut-off at 1\,hr$^{-1}$.  The
CCFs of the filtered lightcurves are shown in the lower panel of
Fig.~\ref{CCFFig}.  On the first night, there is still no prominent
peak in the CCF, in spite of apparent similarities in the lightcurves.
For the second night, the structure at negative lags has been removed,
but the peak remains prominent and definitely appears significant.  We
estimate a delay of the line with respect to the continuum from the
filtered lightcurves of $(70\,\pm40)$\,s, consistent with zero at the
2\,$\sigma$ level.  The error estimate follows Gaskell \& Peterson
(1987) and is defined for uniformly sampled data.  Our data are close
to, but not perfectly, uniform. Hence we believe this error estimate
is not too far wrong.  We also calculated discrete CCFs (Edelson \&
Krolik 1988) to check that the peak was not being pulled towards zero
lag by correlated noise in the line and continuum lightcurves; this
method eliminates correlated noise by only including the contribution
of line and continuum from different spectra.  The discrete CCF
actually yielded a slightly shorter lag of $\sim55$\,s.  Since the
measured delay is not significant at the 3\,$\sigma$ level, and is
below the time resolution used, it is probably not real.  The
3\,$\sigma$ upper limit on the lag is then about 200\,s.  We note that
important short timescales such as the light travel time across the
binary ($\sim30$\,s) are not resolved, so a smaller lag may be
present; higher time-resolution observations would be required to test
this, but would then require an 8--10\,m class telescope.

For the analysis of the line profiles in Section~\ref{ProfileSection},
we divided the lightcurves up into segments, a convenient but
artificial division of continuously varying behaviour.  We identify
the lowest region of each night with the `quiescent state', Q.
Isolated flickering regions, somewhat brighter than Q are labelled F.
Within the flares we separate flare rise (fR), peak (fP) and decline
(fD) segments.  This nomenclature is summarised in Table
\ref{ProfileTable}.

For the two major flares we estimated the peak flare luminosity
emitted in \Halpha\ and the total energy released in this line.  We
assumed that the lowest level reached on each night represents the
base, persistent flux and integrate over the range actually observed.
Since both flares extend beyond our coverage this is a lower limit to
the total energy.  For the first night we estimated a peak flare
luminosity (at 3\,kpc) of $1.4\times10^{32}$\,\ergspersec\ and a total
observed \Halpha\ energy of $\ga9\times10^{35}$\,erg.  For the second
night the values are $1.5\times10^{32}$\,\ergspersec\ and
$\ga9\times10^{35}$\,erg respectively.  The peak luminosities of the
flares are strikingly similar, although the total energies are likely
different, since most of the decline of the flare on the first night
was missed.

\subsection{The high spectral resolution spectra}
\begin{figure*}
\begin{center}
\noindent
\epsfig{width=2.6in,file=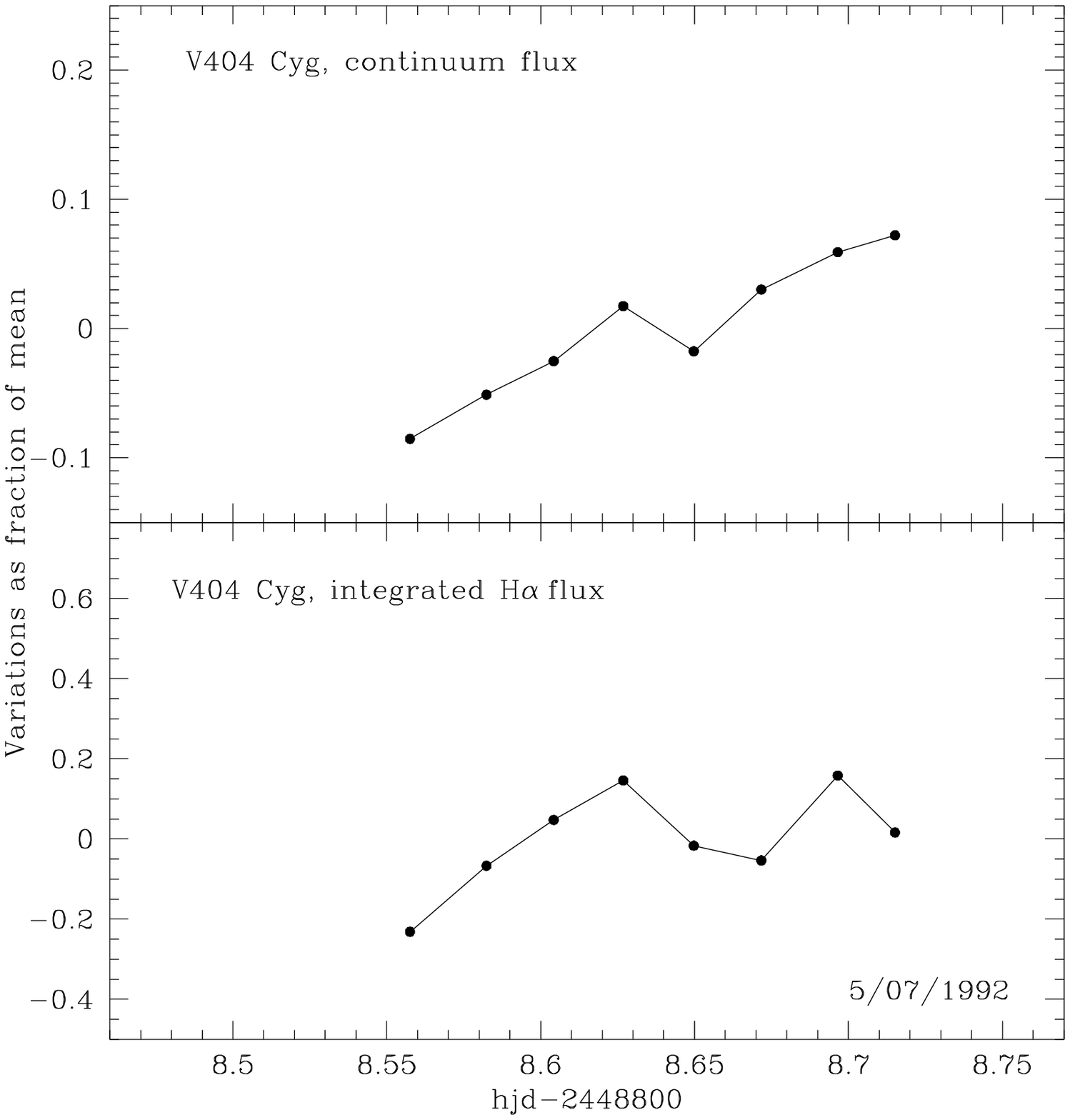}\hspace*{-10mm}
\epsfig{width=2.6in,file=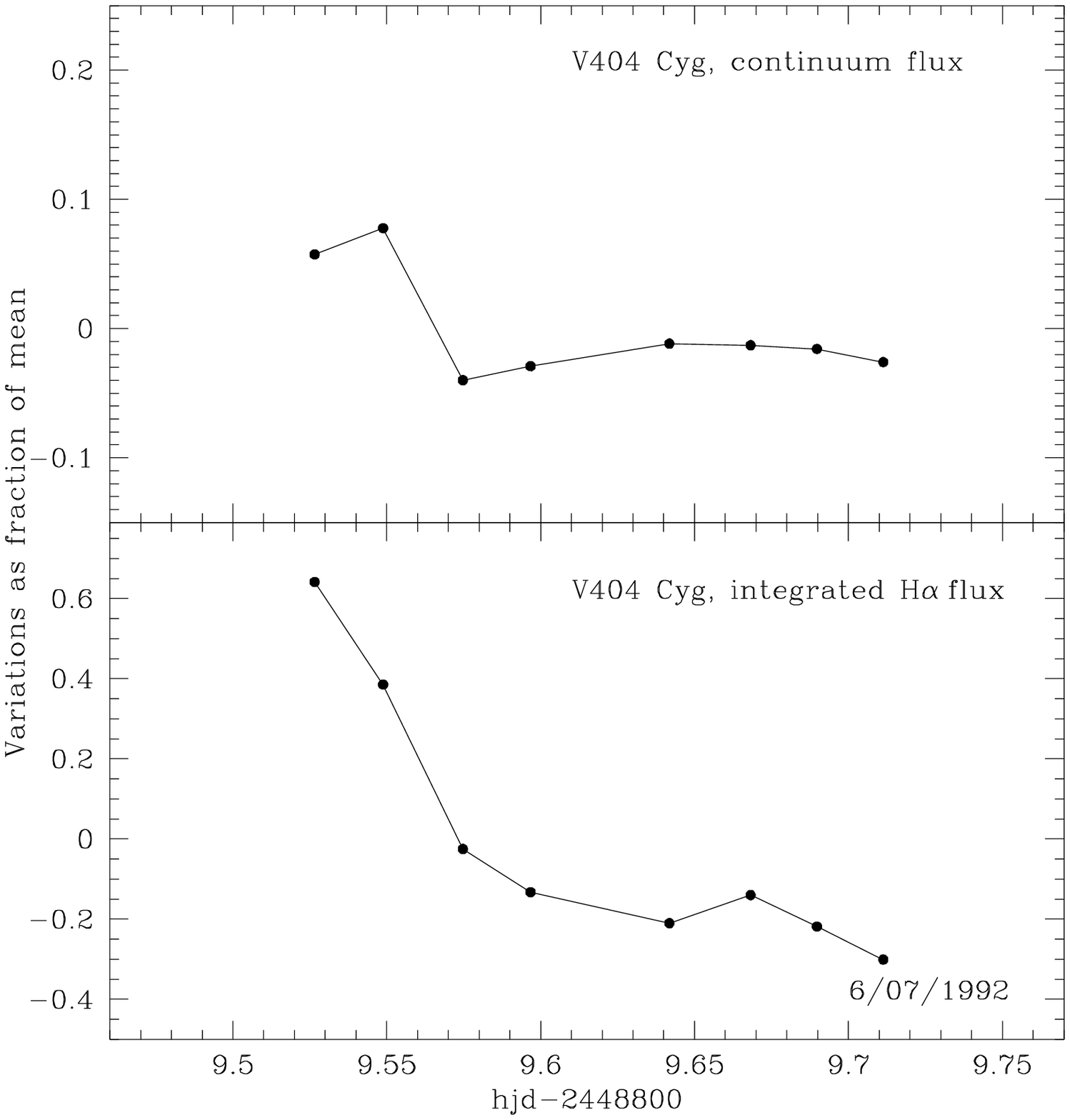}\hspace*{-10mm}
\epsfig{width=2.6in,file=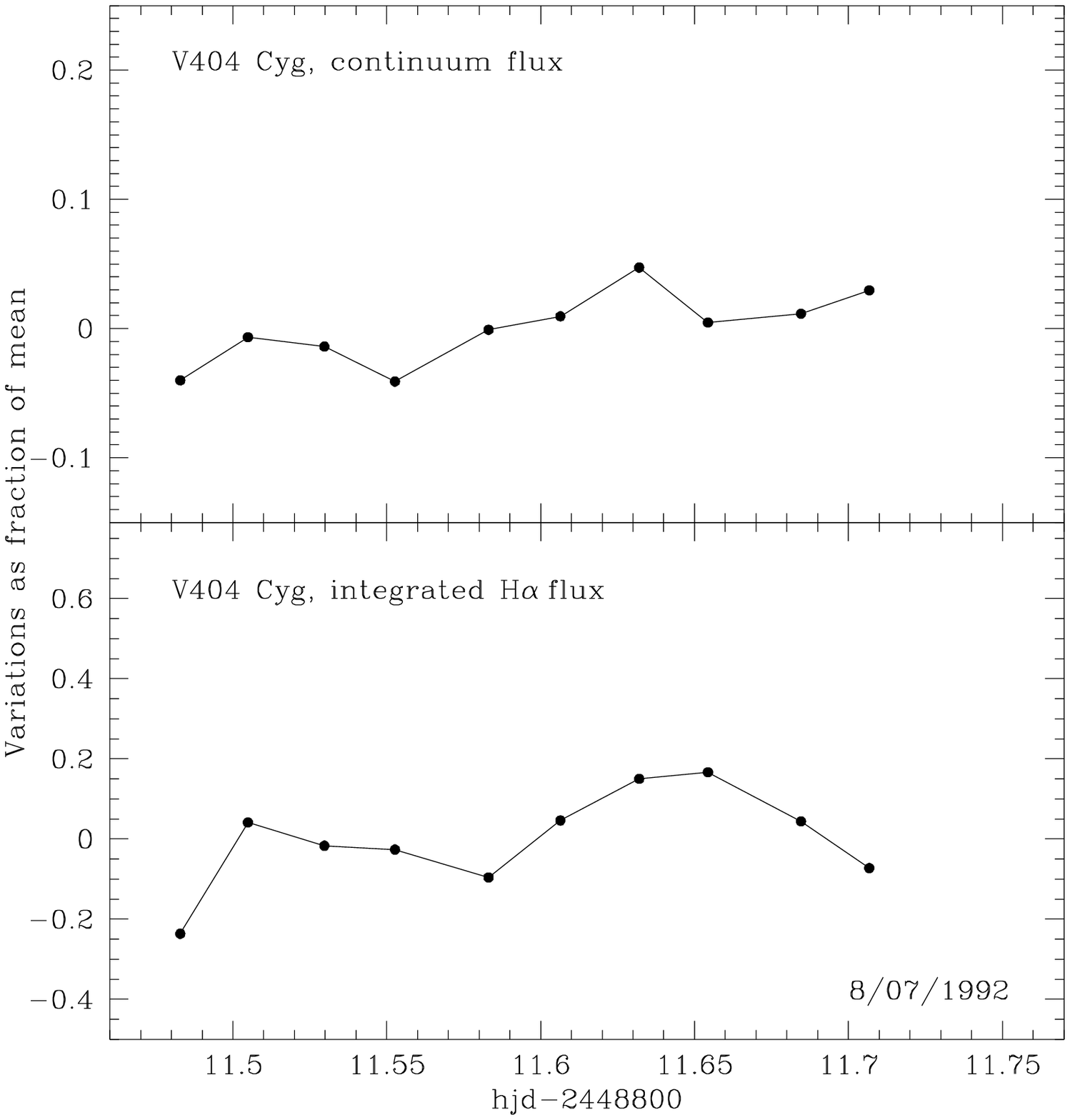}\\
\vspace*{-5mm}
\epsfig{width=2.6in,file=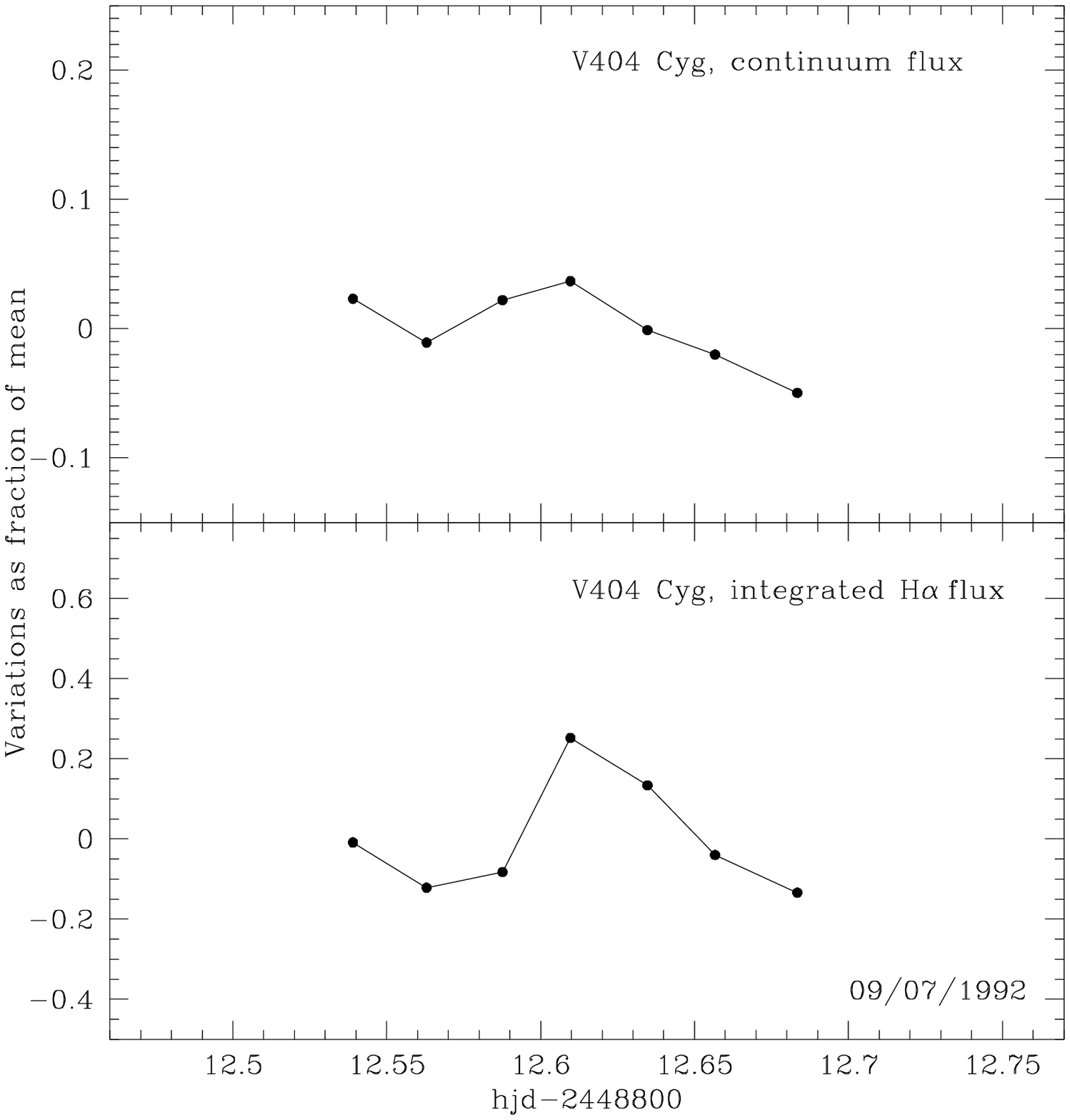}\hspace*{-10mm}
\epsfig{width=2.6in,file=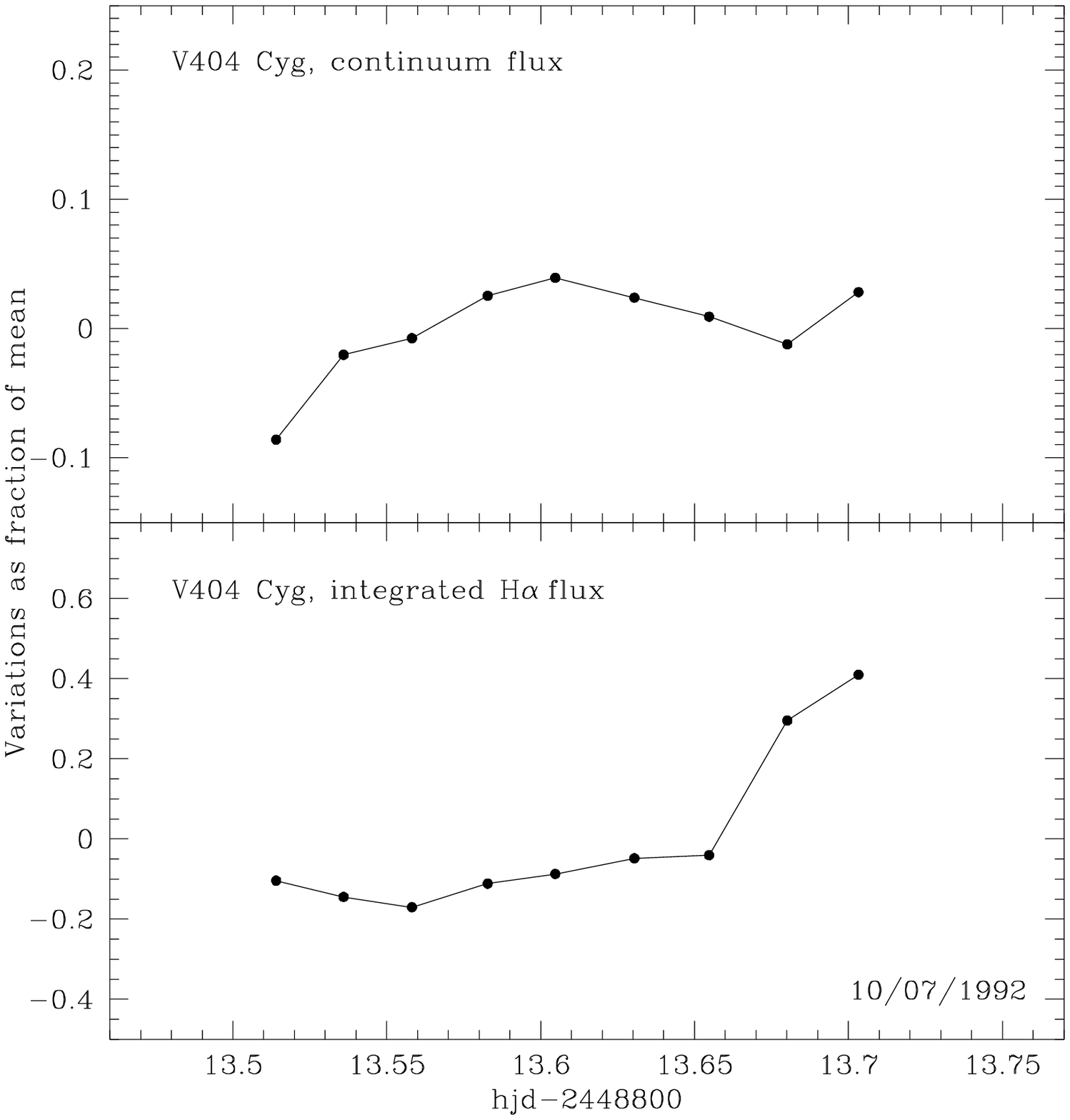}\hspace*{-10mm}
\epsfig{width=2.6in,file=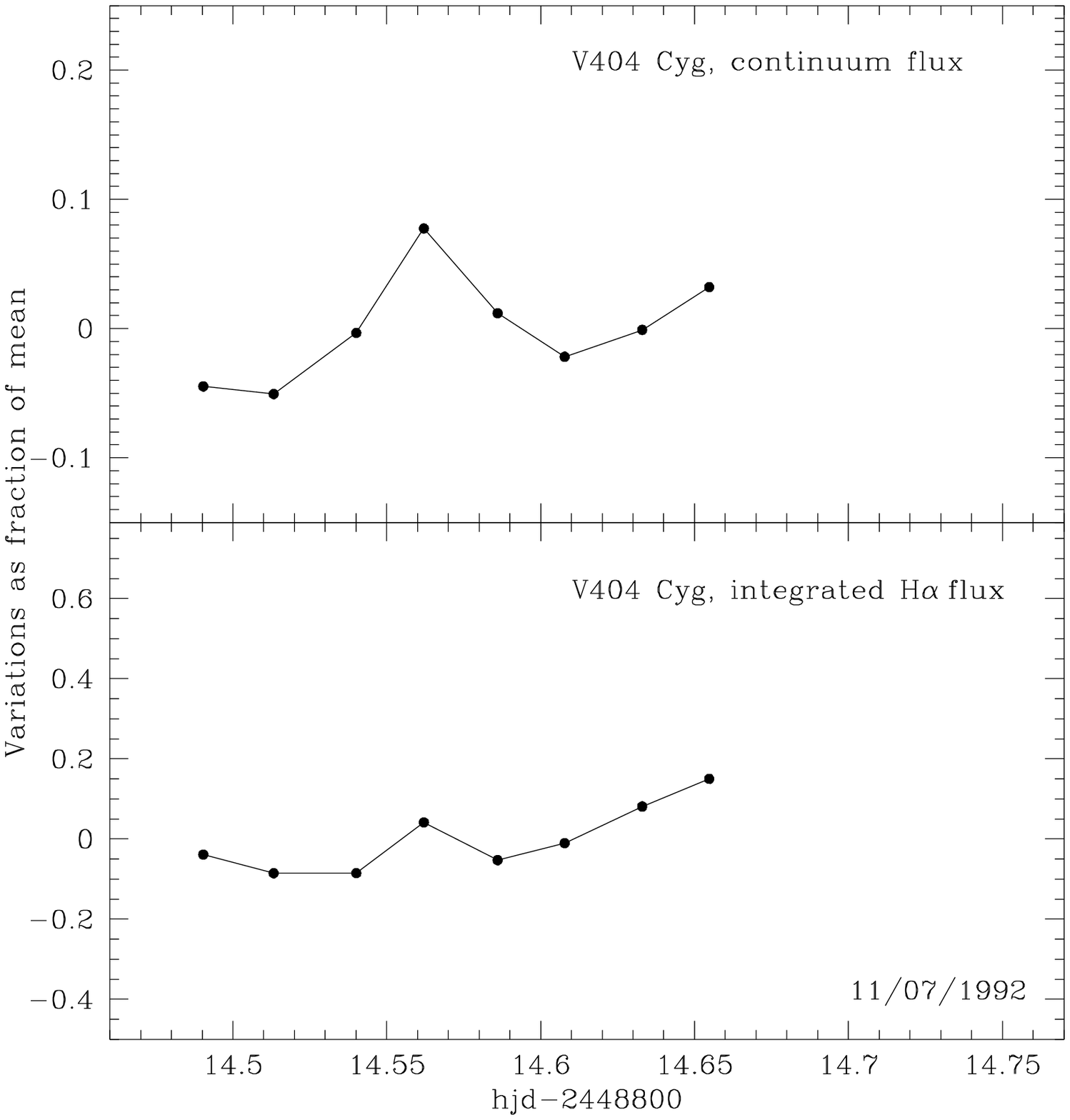}
\caption{Continuum and H$\alpha$ emission lightcurves of \novacyg\ in
1992 July 5--6 and 8--11.  Continuum lightcurves are based on
integrated (unweighted) flux within the bandpass.  Note that the
vertical range of the \Halpha\ plot is much larger than for the
continuum light curves and that the scale is the same as in
Fig.~\ref{LCFig}.}
\label{LCFig1992}
\end{center}
\end{figure*}

Light curves from the 1992 high spectral resolution spectra are shown
in Fig.~\ref{LCFig1992}.  Here we can see that variability is also
dominated by single flares with changes in flux up to a factor of 2 on
time scales of 1--2\,hr, although due to the poorer time resolution,
we cannot resolve any short-term structure.  The largest flares are
seen on the nights of 6, 9 and 11 July and both the amplitude of the
events and the duration is similar to that found in 1999, seven years
later.  
Generally the continuum lightcurve is correlated with the \Halpha\
one, although the amplitude of continuum variations is
$\la50$\,percent of the \Halpha\ variability.

%
%
\section{Spectral Variability}
\label{SpecVarSection}
Lightcurves are a good way to contrast line and continuum variability
but information about velocities is lost.  In this section we explore
several methods to analyse the changes in the line morphology and
constrain the binary region responsible for the \Halpha\ variability.

\subsection{Trailed Spectra}
\begin{figure*}
\begin{center}
\epsfig{angle=90,width=3.4in,file=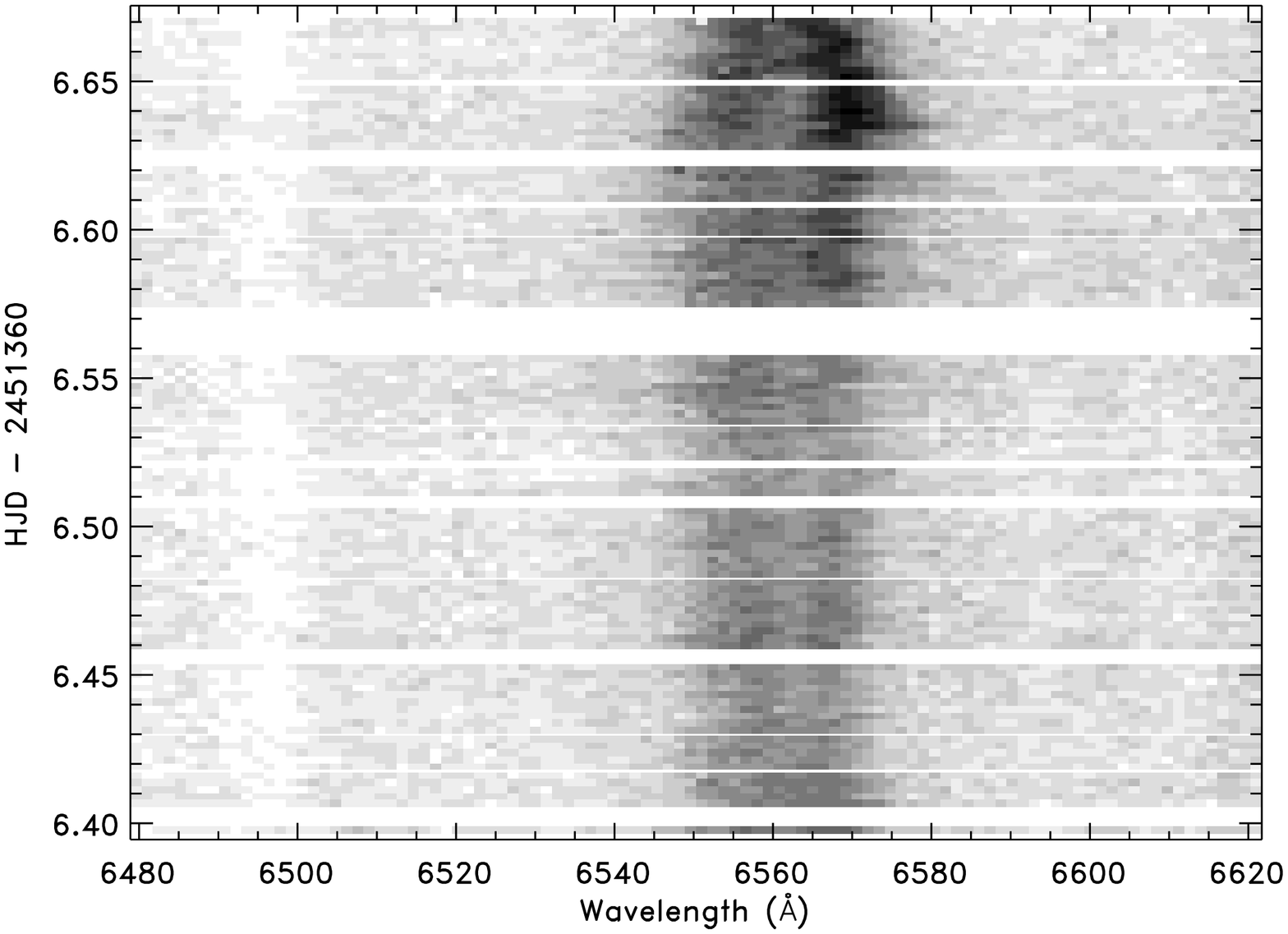}
\epsfig{angle=90,width=3.4in,file=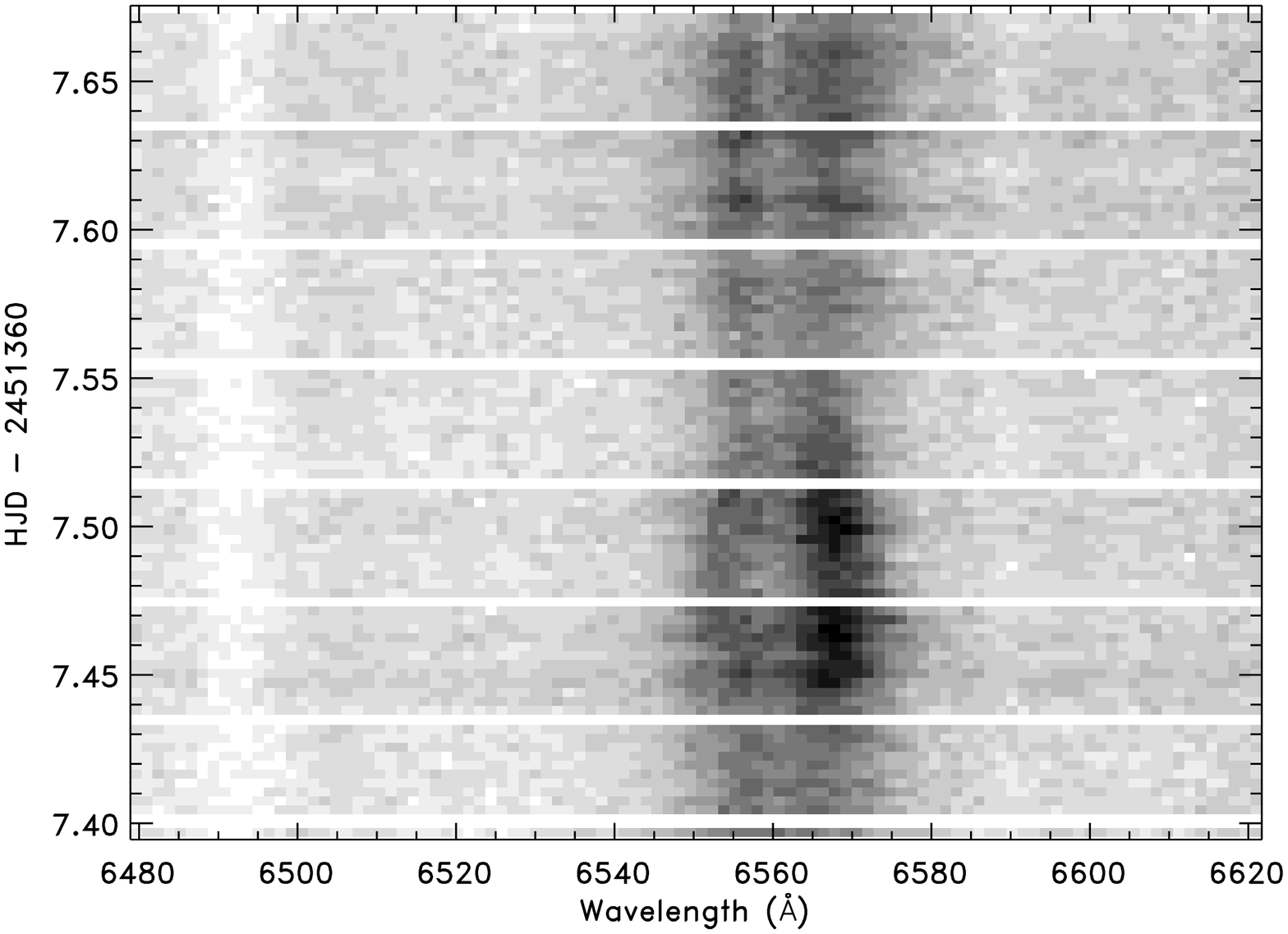}
\caption{Trailed spectra in the vicinity of \Halpha\ from the 1999
observations on July 6--7 (left) and 7--8 (right).  White gaps
indicate when calibration frames were taken.}
\label{TrailFig}
\end{center}
\end{figure*}

\begin{figure*}
\begin{center}
\epsfig{width=2.3in,file=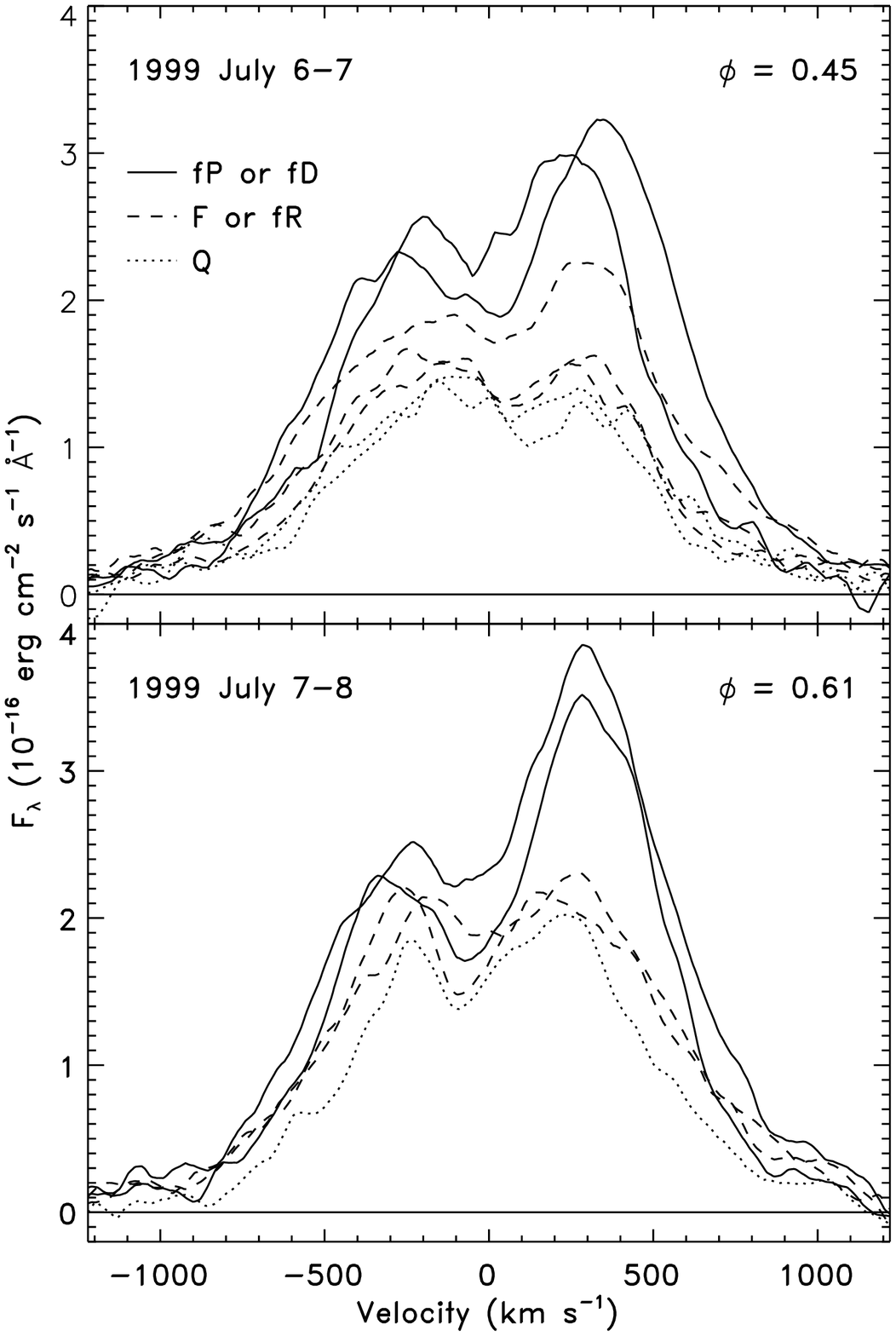}
\epsfig{width=2.3in,file=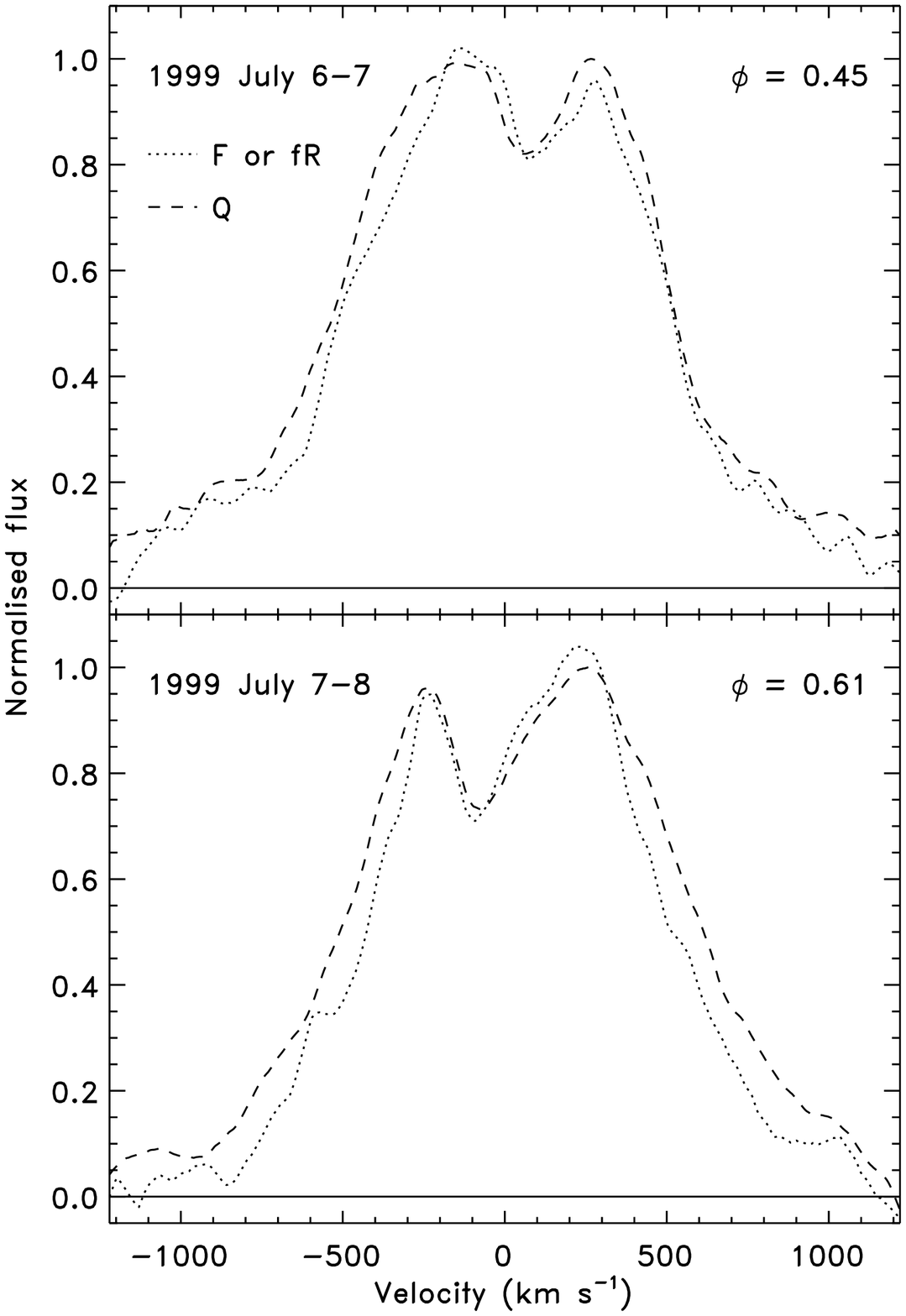}
\epsfig{width=2.3in,file=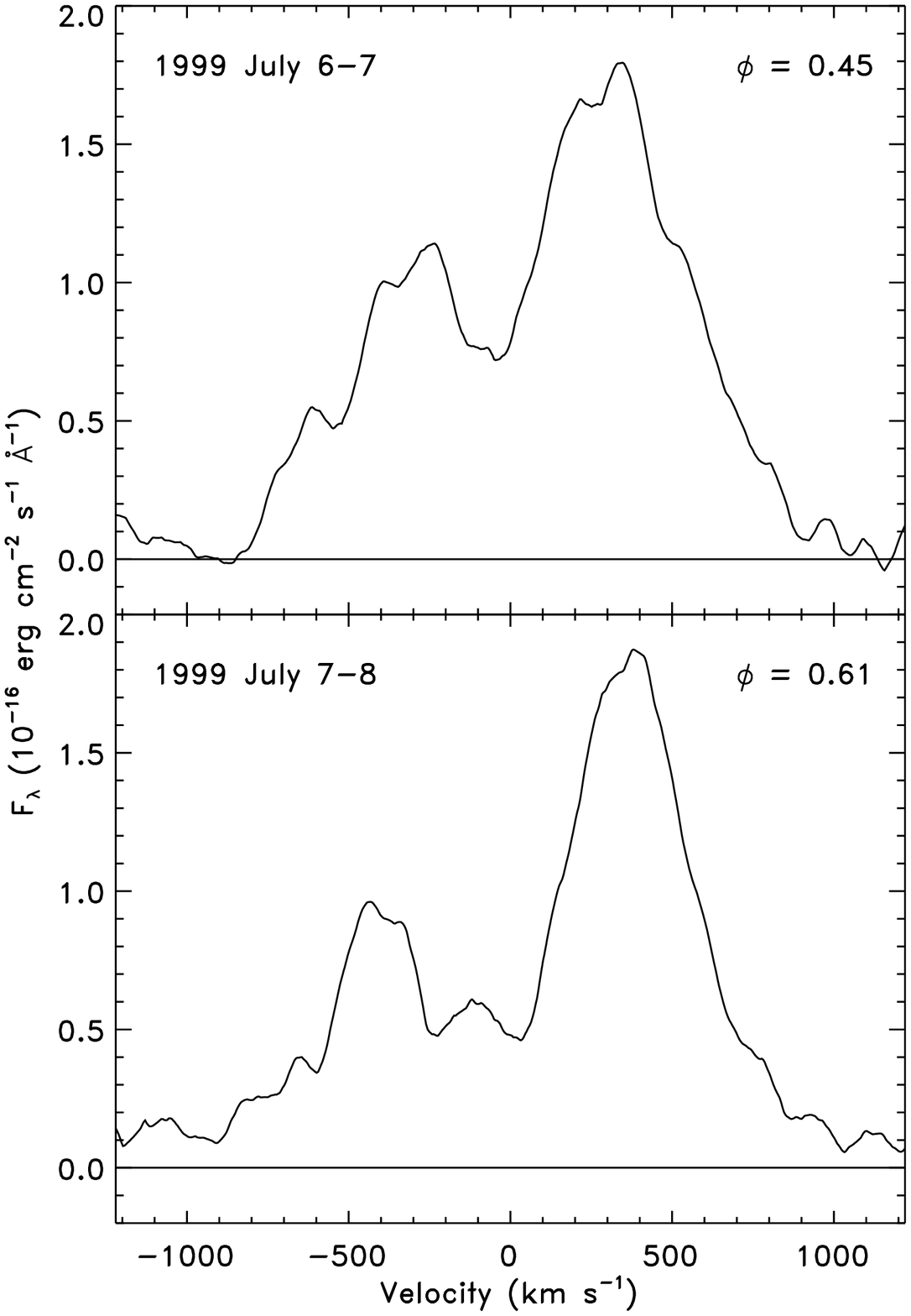}
\caption{Changes in line profiles for the 1999 observations.
Left-hand panel: Calibrated fluxes for all profiles.  Center panel:
F/fR and Q profiles normalised to the same peak flux with all profiles
of the same type from each night have been averaged together.
Right-hand panel: Difference between average fP/fD and Q profiles for
each night.}
\label{ProfileFig}
\end{center}
\end{figure*}

Our trailed spectra from 1999 are presented in Fig.~\ref{TrailFig}.
It can be seen that the line profile varies with the flares, with
strong red peaks seen at flare maxima.  Other than this enhancement of
the red wing relative to the blue, we find no evidence of velocity
changes, or new velocity components or lines appearing during the
flares.  There is an apparent broadening of the profiles in the
trailed spectra during the flares, but this is not real, as can be
seen by the profiles in Fig.~\ref{ProfileFig}; it is due to the
overall brightening of the profile.

\subsection{Line profiles from the high temporal resolution spectra}
\label{ProfileSection}

We have constructed average line profiles from each of the lightcurve
regions indicated in Fig.~\ref{LCFig}.  These are shown in the
left-hand panel of Fig.~\ref{ProfileFig}.  During faint (Q) periods
the \Halpha\ line is largely symmetric and exhibits a `classical'
double-peaked disc profile.  During flickering (F) periods a similar
symmetric disc profile is seen, but it is stronger.  Profiles on the
rise to a flare (fR) are mostly similar, the exception being the
second fR period on the first night (the highest dashed line in Fig.\
\ref{ProfileFig}, just before the peak of the flare.)  In this case, a
mild asymmetry had developed and the profile was clearly developing
towards the fP form (see below).  Excluding this case, F and fR
profiles are indistinguishable.  To test for more subtle differences
between Q and F/fR profiles we combined all of the profiles of each
type (excluding the anomalous fR from the first night) and normalised
the averages to the same total line flux.  These normalised profiles
are plotted in the centre panel of Fig.~\ref{ProfileFig}.  On both
nights there is a tendency for F/fR profiles to be broader than Q
ones.  This suggests that the brightening mainly involves higher
velocity material.  In principal an average profile could also be
broadened by errors in wavelength calibration which, as described in
Section~\ref{SpecRedSection}, is not very reliable.  The rms scatter
in the wavelength corrections was 0.5\,\AA\ and 1.2\,\AA\ on the first
and second nights respectively corresponding to 20 and
50\,km\,s$^{-1}$ at \Halpha.  To explain the observed broadening
requires calibration errors of order 100--150\,km\,s$^{-1}$, however,
so the broadening is likely real.  The spectra at and after the peak
of the flares (fP and fD) are qualitatively very different from those
at lower levels.  On both nights, a strong red peak developed during
the flare; the asymmetry is most pronounced on the second night.  The
blue peak is clearly also enhanced, as can be seen in the difference
profile (fP/fD -- Q) shown in Fig.~\ref{ProfileFig}.
There are changes in the asymmetry during the flares, but these are
different on each night and there is no repeatable difference between
fP and fD profiles.  The velocity of the red peak is
$\sim200-300$\,km\,s$^{-1}$.  The characteristics of profiles in
different phases are summarised in Table \ref{ProfileTable}.

\begin{table}
\caption{Summary of the `states' used to divide up the 1999 data.
These are identified by lightcurve behaviour, but define different
line profiles as well.}
\label{ProfileTable}
\begin{center}
\begin{tabular}{lll}
\hline
Abbrev. & Type & Line profile             \\
\noalign{\smallskip}
Q  & Quiescent     & Weak, symmetric           \\
F  & Flickering    & Medium, symmetric         \\
fR & Flare rise    & Medium, usually symmetric \\
fP & Flare peak    & Strong, enhanced red wing \\
fD & Flare decline & Strong, enhanced red wing \\
\hline
\end{tabular}
\end{center}
\end{table}

We also constructed continuum spectra for each of these periods to
test if the spectrum becomes redder or bluer during the flares.  We
can find no convincing change in continuum colour between Q and fP/fD
periods, although the quality of difference or ratio spectra between
these periods is very poor.

\subsection{The high spectral resolution spectra}

\begin{figure}
\begin{center}
\hspace*{-1.7in}
\epsfig{width=6in,file=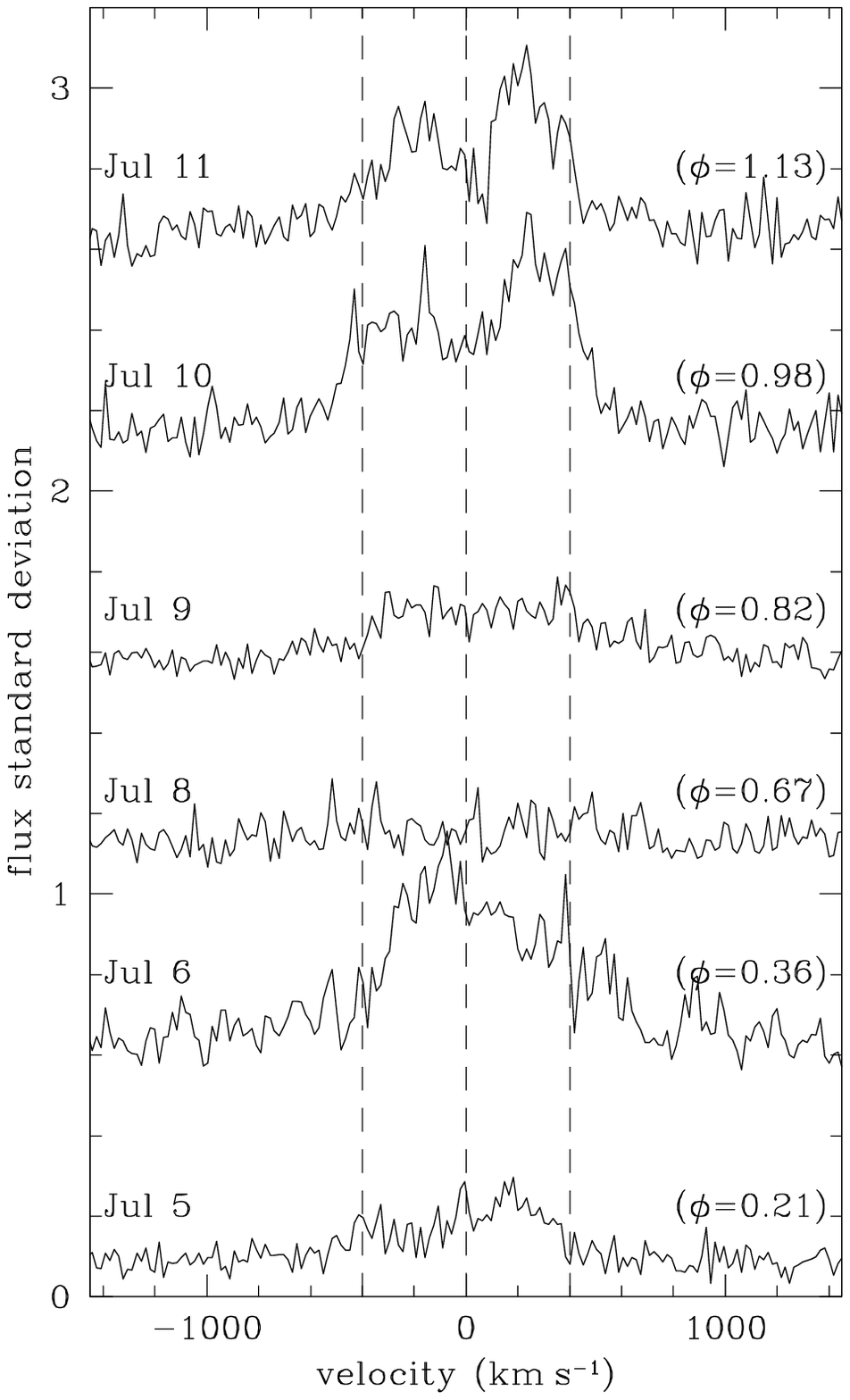}
\vspace*{-20mm}
\caption{RMS spectra for the six nights in the 1992 dataset. Vertical
dashed lines at $\pm400$\,km\,s$^{-1}$mark velocities from the red and
blue cores and wings.}
\label{Devflare92}
\end{center}
\end{figure}

For the 1992 data, the time-resolution is poorer making it hard to
define regions of the lightcurves in the same way as was done in
Section~\ref{ProfileSection}.  These data do, however, benefit from a
higher signal-to-noise ratio and better wavelength calibration.  They
are thus well suited to calculation of the root-mean-sqare (rms)
spectrum, as is commonly employed in AGN studies (see e.g.\ Wandel,
Peterson \& Malkan 1999 for definition).  We find that the line core,
especially the red core (regions confined within
$\pm400$\,km\,s$^{-1}$), is the main contributor to the line
variability (see Fig.~\ref{Devflare92}) on July 5, 10 and 11, peaking
at $\sim+200$\,km\,s$^{-1}$, comparable to the peak velocities in
1999.  On July 6, however, it seems to be the blue core that varies
strongest.  The other two nights do not show pronounced changes in the
symmetry of the lines.  We cannot find any narrow components
associated with the expected velocity of the secondary star or the
hot-spot/gas-stream. The line appears to be more symmetric and
narrower in the low states, as found in the higher time-resolution
spectra discussed above.

%
%
\section{Discussion}
\label{DiscussionSection}
It is clear that significant variability is present in both the
continuum and the \Halpha\ line.  We note that with an average
equivalent width of $\sim20$\,\AA, \Halpha\ will make a small
contribution to $R$ band photometry $\la2$\,percent, so $R$ band
photometric variability should be dominated by the continuum.
Possibly there is more than one kind of variability present, with
different physical origins.  Very prominent are the large flares that
dominate each night's light curve.  These last a few hours each and
occur once per night, making it likely that these can be identified
with individual events of the 6\,hr QPO described by Pavlenko et al.\
(1996).
In addition to these large extended flares, shorter timescale
variability is present with a lower amplitude.  While this may
originate from the same mechanism responsible for the larger flares,
it could also represent a different process.  The timescales involved
in this higher frequency variability are comparable to those found in
shorter period BHXRTs (Haswell 1992; Zurita et al.\ in preparation).
This minute timescale flickering may be typical of quiescent BHXRTs,
but the more dramatic flares could be unique to \novacyg\ and
associated with its longer orbital period, higher black hole
mass, higher X-ray luminosity and/or some other factor.
Alternatively, if the variability timescale scales with orbital
period, then the variations seen in other BHXRTs correspond to the
large flares in \novacyg, and the counterparts of the short-term
variation seen in the latter have yet to be found.

Several variability sites are possible.  An obvious candidate would be
local magnetic flares (reconnection events) within or above the disc.
Other possibilities include instabilities in the accretion stream,
resulting in variability at the stream-impact point, chromospheric
activity on the companion star and instabilities in the inner
accretion flow resulting in variable photoionisation by a central
X-ray source.

\subsection{Kinematic evidence for the flare location}

\begin{figure}
\begin{center}
\epsfig{angle=90,width=3.4in,file=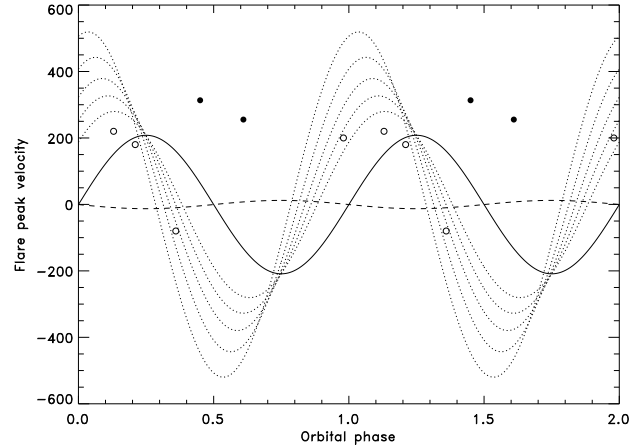}
\caption{Velocities of flare peaks from 1992 (open circles) and 1999
(filled circles).  The solid line shows the radial velocity curve of
the companion star, the dashed line is the curve for the compact
object and the dotted lines are a selection of ballistic stream
velocities at stream impact points corresponding to disc radii of
0.5--0.9 times the lobe radius.  See text for details.}
\label{FlareVelFig}
\end{center}
\end{figure}

An obvious diagnostic of the flare location is the line velocities, as
these should show a radial velocity modulation corresponding to the
flare site.  The asymmetry seen in the flare profiles obviously
complicates matters; this is discussed in Section
\ref{AsymmetrySection}.  We would expect this asymmetry to be applied
relative to the motion of the flaring component, so the underlying
radial velocity modulation should still be apparent.  Fig.\
\ref{FlareVelFig} shows the peak velocities of observed flares
together with several predicted radial velocity curves.  The companion
star's curve is taken from Casares \& Charles (1994).  The compact
object's curve is also derived from this assuming $q=0.06$ (Casares \&
Charles 1994).  For the stream impact point we calculate ballistic
stream velocities corresponding to impacts at 0.5, 0.6, 0.7, 0.8 and
0.9 times the effective lobe radius (Eggleton 1983).  We assume
$q=0.06$, $M_1 = 12$\,M$_{\odot}$ and $i=56^{\circ}$ (Shahbaz et al.\
1994).  These parameters obviously have some uncertainty, but at least
illustrate the range of velocities which should be present.  The
relative constancy of the flare velocities at the range of orbital
phases observed is evidence against the flares originating in the
companion star's corona, and a hotspot origin is similarly unlikely.
Instead the flares appear to originate in or above the disc.  In fact
the lack of scatter in the flare velocities is also difficult to
reconcile with localised flares in the disc; these should be scattered
across its surface, with a spread in peak velocities comparable to the
width of the double peaked profile.  This uniformity of flare
velocity, together with the apparent participation of the whole double
peaked profile in the flare (clearly visible in the difference
profiles in Fig.~\ref{ProfileFig}) argues that the \Halpha\ flare
emission is spread over the {\em whole disc}.  If this is true then
the flares most likely originate in photoionisation by variable
X-rays.  Unfortunately, given the limited number of flares observed,
none of these arguments are statistically conclusive, so we now
examine other evidence to support or reject this interpretation.


\subsection{An active companion star?}

During quiescence the optical emission of BHXRTs is dominated by light
from the secondary; hence the observed variability could be due to
chromospheric flare activity on the companion star.  High levels of
activity can be best maintained via tidal locking in tight binaries,
as occurs in RS CVn systems, so we can expect the same behaviour in
LMXBs.  Comparisons with RS CVn stars have already been made by
Bildsten \& Rutledge (2000) to explain the quiescent X-rays.  They
conclude that the X-ray luminosities seen in most quiescent BHXRTs are
consistent with an origin in coronal activity on the companion star
($L_{\rm X}$ up to $10^{32}$\,\ergspersec\ is seen in RS CVn systems;
Dempsey et al.\ 1993), but explicitly exclude \novacyg\ from this as
its X-ray luminosity is too high.  Lasota (2000), however, goes
further and argues that coronal X-ray emission is expected to be too
faint to wholly explain quiescent X-rays in any BHXRTs.  {\it Chandra}
spectroscopy supports this, indicating that the X-ray spectra of
BHXRTs in general, and V404~Cyg in particular, are harder than those
of stellar coronae (Kong et al., in preparation).  It remains likely
that the companion stars are active and that they do make some
contribution to the X-ray emission, but this probably does not
dominate.  We should then also ask if activity in the chromosphere of
the companion star could be responsible for the optical variability
observed.  Although it has been generally assumed that optical flares
on RS CVn stars are a rare phenomenon, very high levels of activity
have been reported in a few of these systems.  For example, UX Ari is
a double-lined spectroscopic binary with spectral type G5V/K0IV and
6.44\,day orbital period.  It exhibits chromospheric emission
originating from the cooler component.  \Halpha\ flares with peak line
luminosities ranging between $\sim$0.7 and
$\sim1.7\times10^{30}$\,erg\,s$^{-1}$ have been observed (Montes et
al.\ 1996).  These are of the same order of magnitude as the maximum
\Halpha\ luminosity observed in the \Halpha\ flares of other RS CVn
systems such as V711 Tau (Foing et al.\ 1994) and HK Lac (Catalano \&
Frasca 1994).  The dereddened peak luminosities in the \novacyg\
flares are $\sim1.5\times10^{32}$\,\ergspersec\ on 1999 July 6--8, two
orders of magnitude higher than in these RS CVn systems and so
apparently not consistent with an origin in the chromosphere of the
companion star, if it is comparable to an RS CVn star.  This is open
to question, however, as the companion star to V404 Cyg may be a
stripped giant (King 1993).  In this case the internal structure will
be very different to a `normal' RS CVn star with the same orbital
period and spectral type.  Of course, given the line profile of the
flares (Fig.~\ref{ProfileFig}), with an enhancement of both wings, the
kinematics of the flares are also not consistent with an origin on the
companion star, and overall we feel this interpretation can be rejected.

\subsection{Variable photoionisation from the central source?}

The most likely site for the \Halpha\ flares is then the disc.  We
next ask how are they powered?  The most likely possibilities are i)
variations in the central X-ray source producing a variable
photoionisation, or ii) local flares within the disc or disc corona
powering the flares directly.

Support for the variable photoionisation interpretation is provided by
Casares et al.\ (1993) who find \Halpha/\Hbeta/\Hgamma\ ratios
consistent with case B recombination and conclude that the emission
lines are photoionised.  To further test this, we estimate the photon
flux in \Halpha\ and compare with $\it Chandra$ observations reported
by Garcia et al.\ (2001).  The \Halpha\ luminosity estimated in
Section~\ref{SpecSection} is $1.4\times10^{32}$\,\ergspersec,
corresponding to $4.6\times10^{43}$\,phot\,s$^{-1}$.

While a straightforward comparison with the observed $L_{\rm X}$ is
possible it would not be very meaningful.  Any photons above 13.6\,eV
could ionise hydrogen, and the absorption cross-section is highest for
the lowest-energy photons; these will cause ionisations at lower
optical depths where the recombination spectrum is less thermalised.
Further, the X-ray spectrum has photon-index $\Gamma\sim2$ (Garcia et
al.\ 2000), so low-energy photons can be expected to be much more
numerous than their more energetic counterparts.  Consequently, if
photoionisation is important in powering \Halpha\ emission at all, it
will be dominated by the unobservable EUV photons, and to estimate
these we must either adopt a model or extrapolate from the X-ray data.
Extrapolating from the $\it Chandra$ spectrum ($\Gamma\sim2$, $L_{\rm
X} = 4.9\times10^{33}$\,erg\,s$^{-1}$), we expect a luminosity of
$\sim10^{34}$\,erg\,s$^{-1}$ above 13.6\,eV or
$8\times10^{43}$\,phot\,s$^{-1}$.  This is extremely uncertain; for
example modest changes in the photon-index of $\pm0.2$ can change
these numbers by a factor of two.  We feel that at present there is no
definitive model for the high-energy spectra of quiescent BHXRTs that
is more reliable in the EUV than an extrapolation.  Simple ADAF models
do make definite spectral predictions for \novacyg; for example those
presented by Narayan et al.\ (1997) predict an EUV flux above a pure
extrapolation of the X-ray power-law due to an EUV bump from inverse
Compton scattering of synchrotron photons.  Theoretical doubts about
the possible presence of outflows and convection call these models
into question, however (Narayan \& Yi 1994; Igumenshchev, Chen \&
Abramowicz 1996; Blandford \& Begelman 1999).  Models of ADAFs with
winds for \novacyg\ have been considered by Quataert \& Narayan
(1999).  By suppressing the central accretion rate, these increase the
importance of bremsstrahlung radiation from the outer flow relative to
inverse Compton emission and consequently harden the spectrum.  This
gives a poorer fit to observations than a wind-free ADAF, although
this can be rectified by changing assumptions about the microphysics.
In convection dominated accretion flow (CDAF) models emission is also
expected to be dominated by bremsstrahlung radiation from large radii
in the flow and spectral predictions are rather similar to those of
ADAFs with winds (Ball, Narayan \& Quataert 2001).  While the unbroken
power-law of the original ADAF models is in good agreement with the
spectral data, these latter models require some fine tuning of
parameters to achieve a satisfactory fit.
Completely different scenarios exist, for example X-ray emission from
a corona above a cold disc (Nayakshin \& Svensson 2001) but these make
no spectral predictions at all.  Finally, we should remember that the
X-ray flux is known to vary by up to a factor of ten on timescales
longer than the 10\,ks {\it Chandra} observation (Wagner et al.\
1994).  Consequently, the numbers above are only order of magnitude
estimates at best.

Adopting this (with large uncertainties) we require that the disc trap
of order $50$\,percent of the emitted EUV photons, if one \Halpha\
photon is produced per photoionisation.  This is an extremely
simplistic assumption, however, as many of the photoionisations
considered will produce rather energetic photoelectrons.  These could
go on to collisionally ionise another atom, or produce bremstrahlung
photons that produce more photoionisations.  A more secure limit is
provided by comparing the input and output energies; we expect that
the reprocessed emission from a disc element must be less than the
input ionising flux, and so:
\[
L_{\rm H\alpha} = f_1 f_2 L_{\rm X}
\]
where $f_1$ is the fraction of the high-energy emission $L_{\rm X}$
which is input into the disc and $f_2$ is the fraction of the input
energy emitted in \Halpha.  $f_1$ depends on both the geometry and the
local conditions, as any photons which are Compton scattered off the
disc will escape without causing a photoionisation.  We can estimate
an upper limit on $f_2$ in the case B recombination limit (supported
by observations of \novacyg; Casares et al.\ 1993).  In this case all
Lyman line and continuum emission is assumed to be reabsorbed and the
\Halpha\ emission is 20--30\,percent of the total Balmer and Paschen
line and continuum emission for temperatures of 5000--20000\,K
(Osterbrock 1987).  Hence $f_2\la0.3$.  In reality $f_2$ could be
significantly below this limit, as energetic photoelectrons may lose
energy by other processes besides collisional ionisation: scattering
off other electrons and bremsstrahlung may both be significant and
will tend to thermalise the input energy.  This problem will be more
severe for the higher energy photons which will tend to deposit energy
at higher optical depths within the disc and ultimately give rise to
thermal continuum emission.  For $f_2\la0.3$ we then require $f_1 \ga
0.05$, although due to the uncertainty in the incident ionising flux
this number is rather uncertain.

In principal, photoionisation could occur on the companion star or the
disc.  For \novacyg, however, the mass ratio is small ($q=0.06$;
Casares \& Charles 1994) and the companion star will absorb $\la
1$\,percent of the emitted flux from near the central source.  For
the disc, the calculation is straightforward for the simplest ADAF
models with emission from close to the black hole.  For an isotropic
point source illuminating a thin, but somewhat flared disc we expect a
fraction $\sim h/r$ to be absorbed; hence we require $h/r \ga0.05$ for
an ADAF radiating from the centre.  ADAFs with winds or CDAFs,
however, are expected to emit most of their radiation from larger
radii and hence can more efficiently irradiate the disc.  The
difference is probably not huge, however, perhaps a factor of a few
higher.  Unfortunately, given the large uncertainties in extrapolating
the X-ray spectrum into the EUV region and the variability in the
X-ray flux, none of these possibilities can be ruled out and it
appears that it is energetically possible for all of the advective
models considered to produce sufficient photoionisation on the disc to
account for the observed \Halpha\ flares, provided that not too much
of the input energy is thermalised at large optical depths.  Equally
if X-ray emission originates from a corona above the disc (Nayakshin
\& Svensson 2001) then a large fraction ($\sim 50$\,percent) will be
incident upon the surface, so this model is also plausible, even with
emission of a significant fraction of the energy in the continuum.


\subsection{Local magnetic reconnection events?}

An alternative interpretation of the flares would be that they
originate in local magnetic reconnection events in or above the disc.
It is widely believed that a dynamo mechanism operates in accretion
discs, driven by the strong shear produced by differential rotation
(see e.g.\ Tout \& Pringle 1992).
There is limited observational evidence for this, but it is known that
Balmer line emissivity in quiescent dwarf novae seems to follow the
same power law with rotation rate ($\propto R^{-3/2}\propto\Omega_{\rm
Kep}$) as in chromospherically active stars (Horne \& Saar 1991).
Consequently it has been proposed that the mechanism powering the
\Halpha\ emission lines is the same in both active stars and discs, a
dynamo.  The effect of the strong shear is that regions of oppositely
directed magnetic fields can develop within the disc (Tajima \& Gilden
1987).  These fields can reconnect explosively (Haswell, Tajima \&
Sakai 1992) giving rise to a flare.  We also expect flux tubes to rise
from the accretion disc surface into its corona because of the
magnetic buoyancy (Parker) instability as occurs in chromospherically
active stars, so reconnection and flaring could originate there.

In solar flares, it is commonly accepted that every \Halpha\ flare is
accompanied by a soft X-ray event, hence the effects of X-ray
irradiation have to be taken into account and the ionisation and
excitation of hydrogen together with the thermalisation of the
background electrons will be two important processes.  Simultaneous
\Halpha/soft X-ray (0.05--2\,keV) observations of solar flares show a
\Halpha\ to soft X-ray energy ratio of $\sim$5\%
\cite{Acton:1982}. The same ratio has been observed in the
chromospherically active dMe star Gl644AB \cite{Doyle:1987}, where the
total \Halpha\ flare energy is 6\,percent of the X-ray energy in the
the 0.05--2\,keV range.  Although in \novacyg\ no simultaneous
observations are available, a factor of $\sim$2 variability in the
X-ray flux has been observed \cite{Garcia:2000a}, so the peak flare
luminosity is of the same order of magnitude as the average
luminosity, assuming similar spectra.  We can estimate the total
energy in the 0.05--2\,keV range by extrapolating the observed X-ray
power-law ($\Gamma\sim2$) to be $\sim 6 \times
10^{33}$\,\ergspersec. The observed luminosity in \Halpha\ is $\sim
1.5 \times 10^{32}$\ergspersec, that is $\sim$2.5\,percent of the
X-ray energy.  This is a similar order of magnitude to that seen in
solar and stellar flares, although obviously very uncertain since a
large extrapolation of the X-ray spectrum is involved.  The X-ray
observations could therefore be consistent with a magnetic
reconnection scenario, assuming that the ratio of \Halpha\ to X-ray
emission does not radically rescale for flares in or above a disc.
Given the uncertainties, an additional X-ray component from close to
the black hole is certainly still possible even with this
interpretation of the variability.

A major problem for this model is that the timescales involved, with
flare rise times of a few hours, are rather short.  The line profile
enhancements and the large amplitude of the flares suggest that these
are not localised events but that the whole disc brightens on this
timescale.  Given the long period (6.5\,d) of V404~Cyg it is hard to
see how this could happen, as viscous timescales in the disc will be
much longer than this, and hence the whole disc should not be coupled
on this timescale.


\subsection{Line profile variations}
\label{AsymmetrySection}

The development of asymmetry in the line profile is a further
interesting issue, especially as it is usually the red wing that is
affected; only one night, 1992 July 6 shows preferential enhancement
of the blue wing.  In addition to the peaks summarised in Fig.\
\ref{FlareVelFig} the single night of spectroscopy from 1990 July 20
presented by Casares \& Charles (1992) shows a red flare at orbital
phase $\sim0.0$ and in 1991, strong red wings were seen on August 8
and 9, corresponding to phases $\sim0.05$ and $\sim0.20$ (Casares et
al.\ 1993).
This suggests that the enhancement in the red wing is not dependent on
orbital phase, so does not indicate kinematics of localised flares.
It is also notable that it only occurs when the line is strongest;
during fP, fD and high fR periods in the nomenclature of
Section~\ref{LightcurveSection}.  During F and low fR periods, the
line profile is enhanced, but symmetrically.

\begin{figure}
\begin{center}
\epsfig{width=2.5in,file=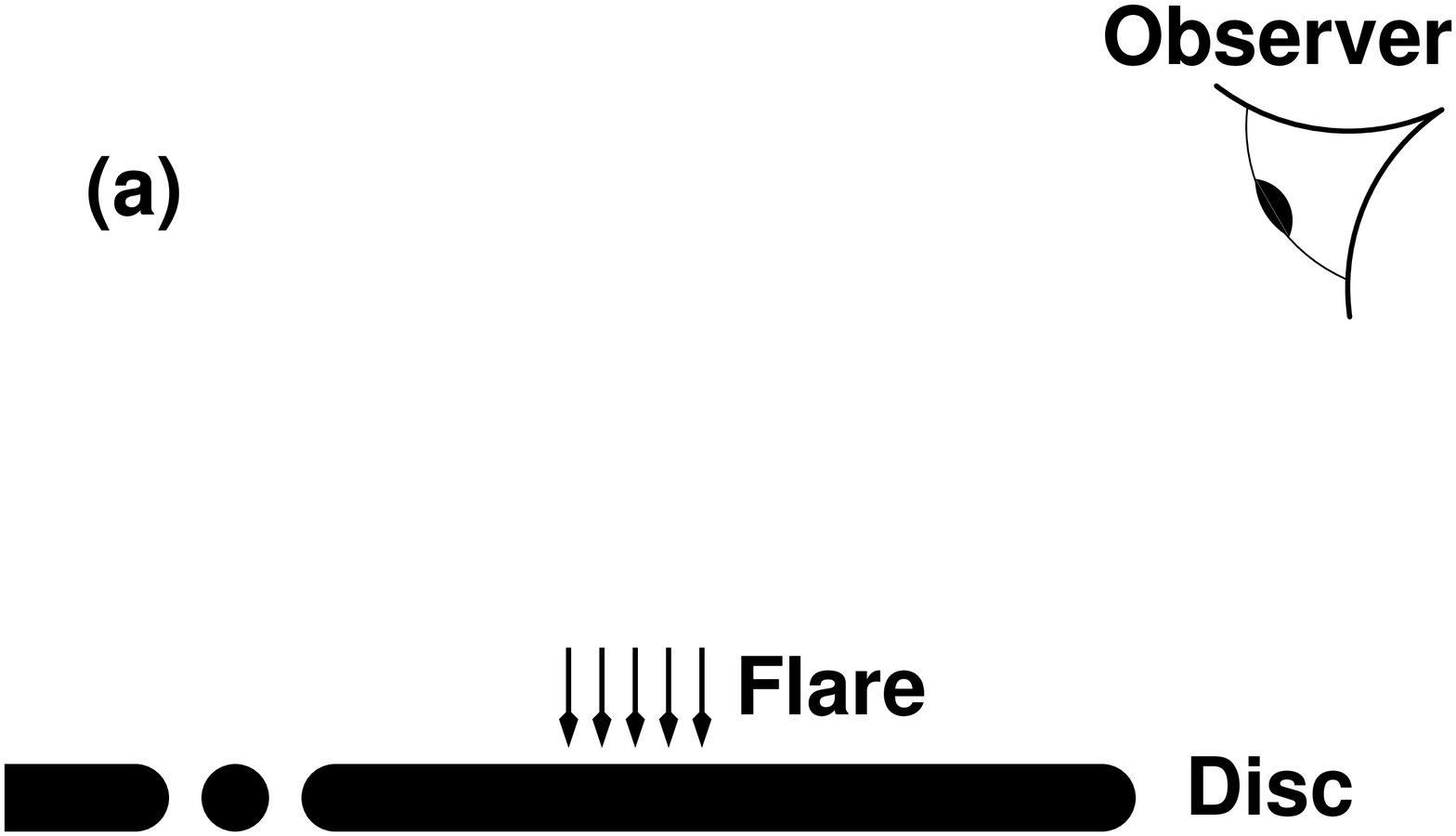}\\
\vspace*{5mm}
\epsfig{width=2.5in,file=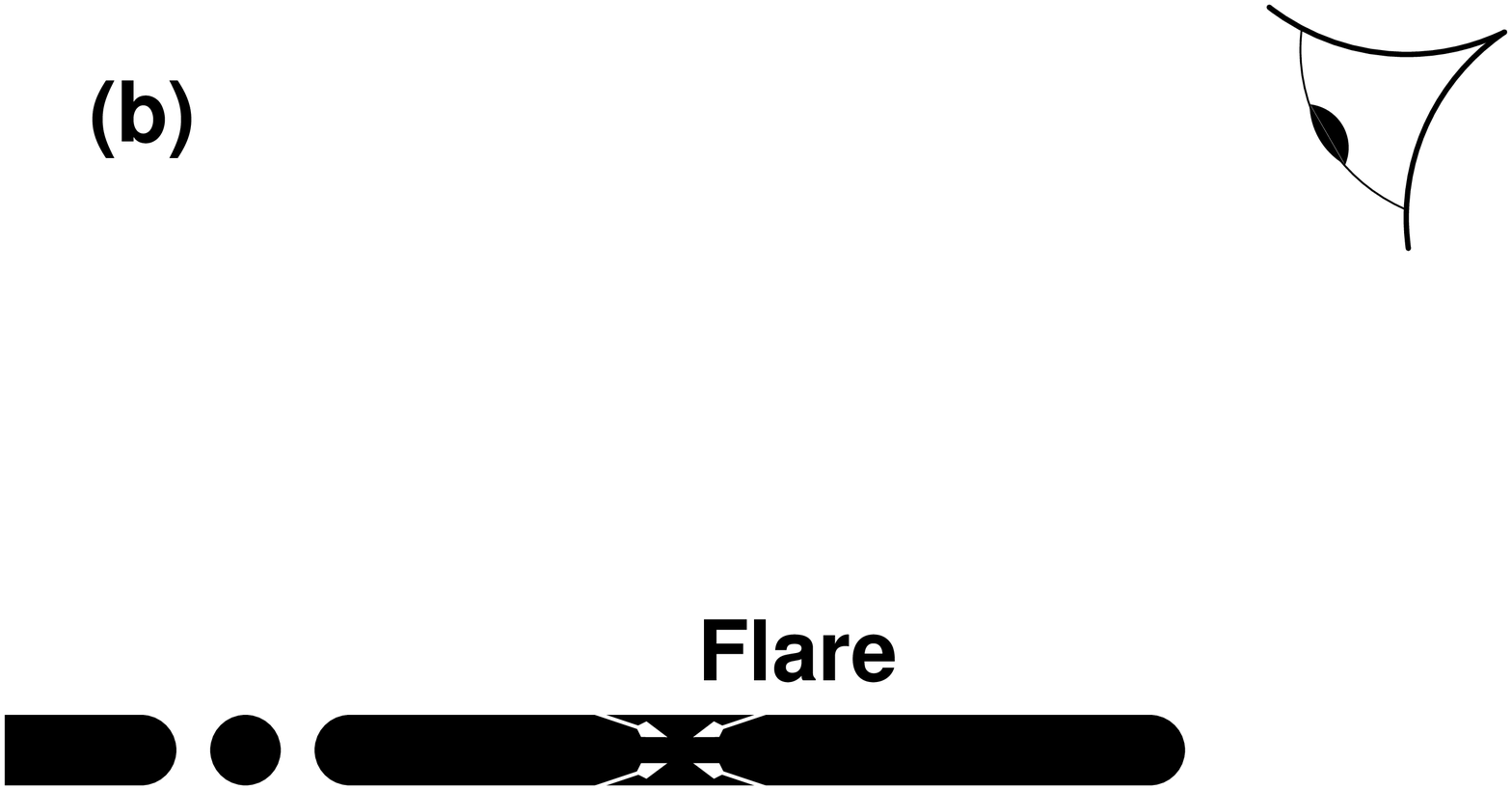}\\
\vspace*{5mm}
\epsfig{width=2.5in,file=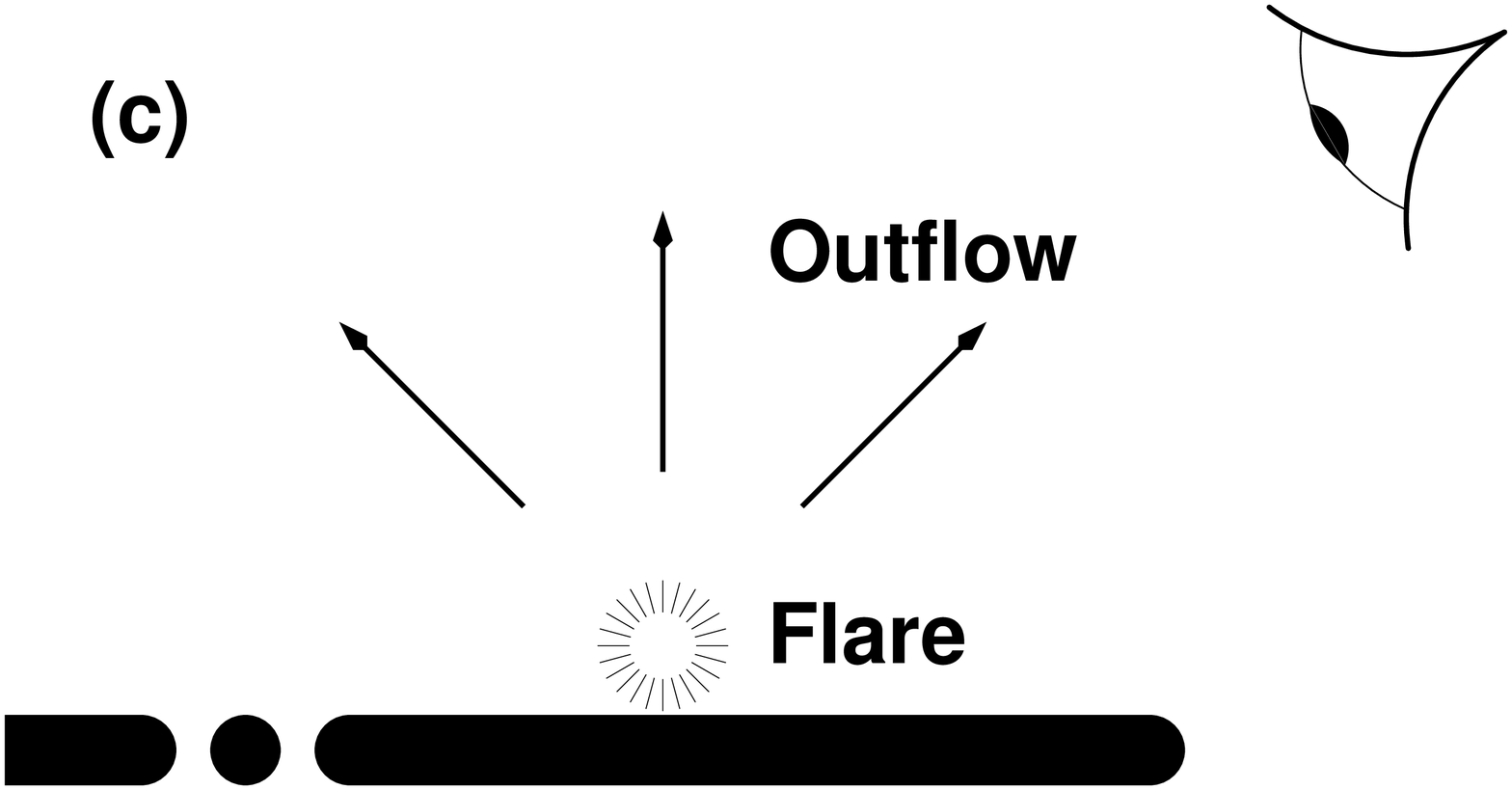}\\
\vspace*{5mm}
\epsfig{width=2.5in,file=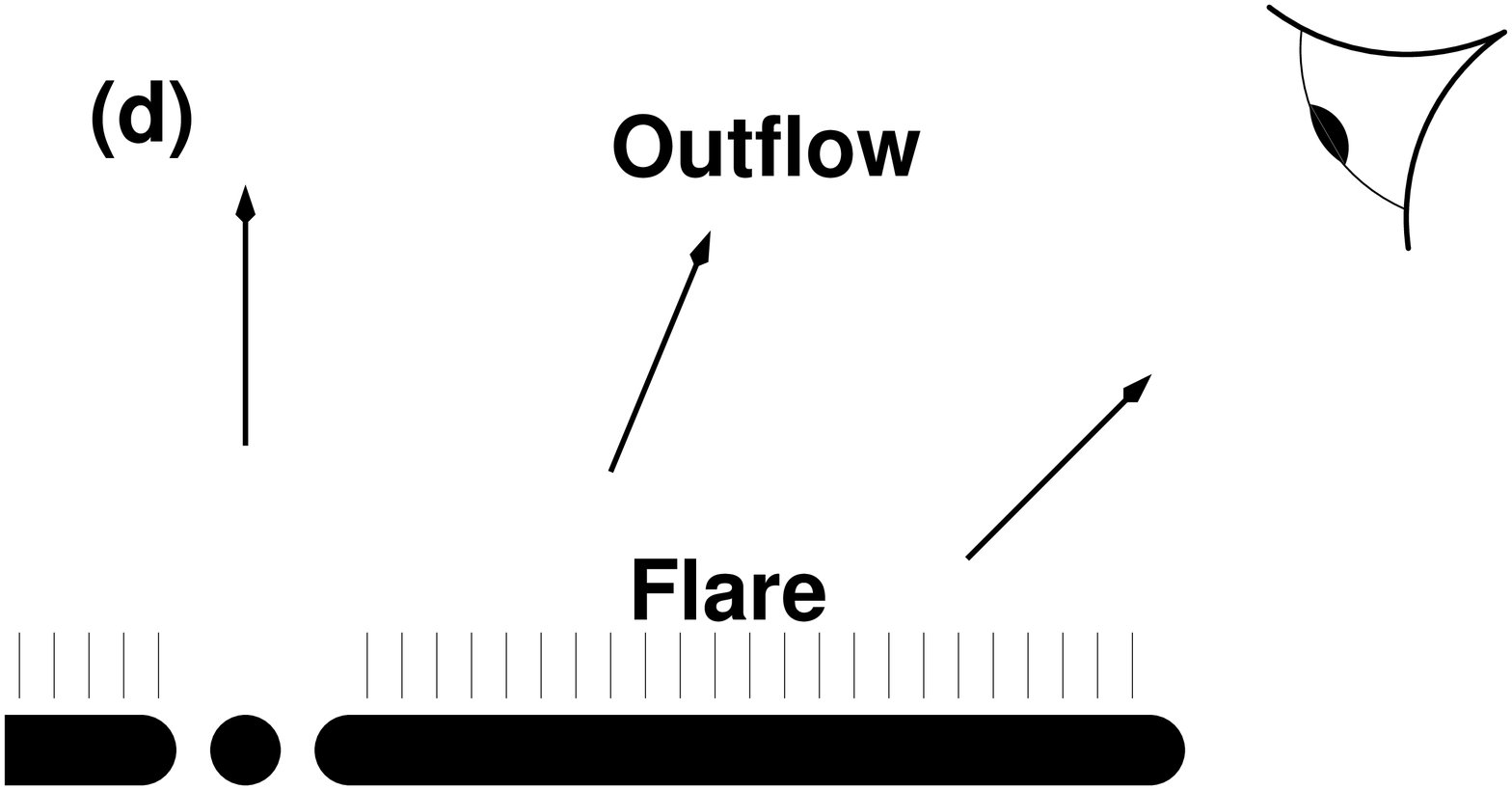}
\caption{Possible geometries giving rise to a redward asymmetry during
flares.  {\bf a)} A flare in the disc corona collapsing downwards onto
the disc.  The emitting material is redshifted.  {\bf b)} A flare
within the disc is accompanied by motion towards the reconnection
point.  The material nearest the observer is redshifted.  {\bf c)} A
flare is accompanied by an outflow.  The portion of the outflow
between the observer and the flare is blueshifted so absorbs the blue
wing of the flare profile.  The outflow could originate from a
reconnection event or be associated with a flare of the central
source. {\bf d)} The line emission from the flare is spread over the
whole disc.  An outflow, either from the central source or the disc,
partially absorbs the blue wing.}
\label{FlareGeometryFig}
\end{center}
\end{figure}

It has long been known that solar flares commonly exhibit a red
asymmetry in the \Halpha\ profile (Ellison 1943; see \u{S}vestka 1962
for a more comprehensive dataset).  The asymmetry abruptly increases
at the onset of the flare, attains a maximum velocity of
40--100\,km\,s$^{-1}$ and decreases before the \Halpha\ line reaches
its maximum (Ichimoto \& Kurokawa 1984).  Early papers interpreted it
as due to absorption of the blue wing of the emission profile, but
Ichimoto \& Kurokawa (1984) associate it with a downward motion in the
flare chromospheric region.  The analogue for an accretion disc would
be a flare above the disc accompanied by a collapse downwards (Fig.\
\ref{FlareGeometryFig}a).  The velocities we observe in \novacyg\ are
larger than in solar flares, but these likely will rescale in a very
different environment.  There is a qualitative difference, however, in
that our asymmetry persists well after the flare peak, and appears to
be more pronounced on the decline than on the rise.

Another mechanism emerges from the model of Haswell, Tajima \& Sakai
(1992).  This model predicts explosive reconnection of wound-up
magnetic fields, accompanied by a bulk motion of plasma towards the
reconnection point.  In this case material closest to the observer
will be red shifted and emission is expected over a range of
velocities (Fig.\ \ref{FlareGeometryFig}b).  The temporal properties
might be expected to be similar to solar flares, however, with the
asymmetry most prominent on the flare rise; this is not supported by
our observations.

Strong red peaks are actually not uncommon in BHXRTs, but are usually
a feature of the outburst.  Wu et al.\ (2001) have discussed the
formation of lines in X-ray binaries and suggest that the asymmetric
profiles seen in outburst are a consequence of absorption by a
relatively slow disc wind driven by external irradiation.  These
asymmetric profiles are often accompanied by broad absorption troughs
which are attributed to absorption by higher velocity material.  In
one case, GRO\,J1655--40, we may also have seen P~Cygni profiles in
the ultraviolet, also indicative of a disc wind (Hynes et al.\ 1998).
In the case of \novacyg, we only see the asymmetric profile without
the broad absorption component.  A plausible interpretation of this
would be that a disc wind, or some other outflow, can be formed during
strong flares.  This could originate either in a flare in the central
source which photoionises the disc, or in outward moving material from
a reconnection event.  In these flares, the whole profile is enhanced,
but the blue wing is absorbed by approaching material (Fig.\
\ref{FlareGeometryFig}c,d).  No higher velocity ejecta are present, so
there is no broad absorption component.  Given that \novacyg\ remains
a variable radio source in quiescence, it is also possible that there
may be connection between such ejecta and the radio variability, with
continued ejection of matter driving the radio variability.
\subsection{The quasi-periodicity}
\label{QPOSection}
The final question to be considered is the origin of the 6\,hr
quasi-periodicity of the flaring (Pavlenko et al.\ 1996).  This was
clear in earlier data, but less conclusive in 1999, although it may
still be present.  If flares originate in local magnetic reconnection
events it is very hard to envisage a mechanism which would make events
in different parts of the disc coordinate on a 6\,hr timescale.  

Esin et al.\ (1997) suggested that interactions between an ADAF and
the outer thin disc might produce quasi-periodic variations with a
characteristic timescale of a multiple of the Keplerian period at the
transition radius.  If we identify 6\,hr with the Keplerian period at
the transition radius, $r_{\rm tr}$, then for a 12\,M$_{\odot}$ black
hole this corresponds to $\log r_{\rm tr} \sim 4.8$ in units of the
Schwarzschild radius.  This is much larger than assumed in ADAF models
and indeed the high velocity wings of \Halpha\ rule out such a large
transition radius ($\log r_{\rm tr} < 4.4$; Narayan et al.\ 1997).
There remains the possibility that the QPO period is a multiple (2 or
3) of the Keplerian period, although why this should be the case is
not clear.  Alternatively, 6\,hr could be some instability timescale
intrinsic to the inner flow.
%
%
\section{Conclusion}
We have found significant spectral variability in the BHXRT \novacyg\
in quiescence; most nights exhibit some variability.  The variations
are most dramatic in \halpha, (and presumably other emission lines as
well), with changes of nearly a factor of two in the line strength
over a few hours.  Similar behaviour is seen in the continuum and this
is usually correlated with line changes.  The continuum variations
show smaller amplitude, as expected if the dominant continuum source,
the companion star, does not vary.  The variability can be
phenomenologically divided into two types.  Low level flickering
involves changes in the line flux, without dramatic changes in the
profile shape.  Large flares seen approximately once per night, and
may be associated with the $\sim6$\,hr QPO, involve the development
of pronounced asymmetry in the line profile with the red wing
strongest.  

We have considered several interpretations of the variability.
Chromospheric activity is seen on other rapidly rotating late-type
stars and this gives rise to \halpha\ flares.  The energy released
by the flares in \novacyg, however, is two orders of magnitude higher
than in such systems, and the line profile changes are hard to
reconcile with this interpretation so we consider it unlikely.
Variability could originate from variations in the mass transfer rate
from the companion star and be manifested by flickering at the stream
impact point.  The kinematics of the observed flares do not appear
consistent with this either, however.

Variability could originate locally within the disc due to magnetic
reconnection events, and may involve photoionisation by X-ray flares
in the corona immediately above the disc.  To explain the
participation of the whole line profile in the events requires a large
spread in velocity of material involved in the flare, but this would
be expected from the explosive reconnection model of Haswell et al.\
(1992).  In this interpretation the redward asymmetry could indicate
either motion of the flaring material away from us or partial
absorption by material moving towards us.  It is unclear, however,
why the flare peak velocities do not vary according to where in the
disc the flare originates.


Instead the observed flare kinematics suggest that: i) the center of
light of the flares does not move significantly over an orbital cycle
and ii) the profile of the flare is quite similar to the disc profile,
except for the stronger red wing.  The most obvious interpretation of
is that the whole disc, centred on the slowly moving compact object
($K_1 \la 15$\,km\,s$^{-1}$), participates in the flare.  It is hard
to see how a magnetic reconnection event could affect the whole disc
in this way, so it is most likely that the flares are induced by
variable photoionisation from the X-ray source.  We have compared the
luminosity of the flares with the expected photoionising flux and find
that this interpretation is plausible; all advective-type models
(ADAFs, ADAFs with winds, CDAFs) could produce the observed \halpha\
emission, although models which predict most of their emission from
large radii in the flow can more efficiently irradiate the disc.  In
this interpretation the redward asymmetry of the flare profiles can
best be explained by absorption of the blue wing, either by an
irradiation-driven disc wind or an outflow from the central source.
The correlation between the line and continuum which we observe is
expected for ADAF models which predict that the optical accretion
continuum is self-absorbed synchrotron from near the compact object.
In CDAF models, continuum emission is attributed to the outer disc,
but it could then result from reprocessed irradiation and so be
correlated with variations in photoionised lines.  The mechanism
responsible for the 6\,hr characteristic timescale in the flaring
remains unidentified.



If this interpretation is correct then \Halpha\ flares should be
accompanied by correlated X-ray flares; this can readily be tested
with simultaneous observations.  If such a correlation is not seen
then a local flare hypothesis becomes more likely.  A possible test of
the magnetic recombination flare hypothesis, would be to measure
linear polarisation in the \Halpha\ profiles during the impulsive
phase of the flares since \Halpha\ and \Hbeta\ have been
found to be linearly polarised in solar flares \cite{Firstova:1996},
with a degree of polarisation $\sim$20\,percent.
%
%
\section*{Acknowledgements}
RIH would like to thank Kinwah Wu, Mike Garcia and members of the
Astronomical Institute `Anton Pannekoek' for constructive discussion
on the interpretation of these results.  RIH, CAH and PAC acknowledge
support from grant F/00-180/A from the Leverhulme Trust.  The William
Herschel and Jacobus Kapteyn Telescopes are operated on the island of
La Palma by the Isaac Newton Group in the Spanish Observatorio del
Roque de los Muchachos of the Instituto de Astrof\'\i{}sica de
Canarias.  This research has made use of the SIMBAD database, operated
at CDS, Strasbourg, France and the NASA Astrophysics Data System
Abstract Service.
%
%

%
\end{document}